%% file: susy.tex
\documentclass[12pt,numbers,sort&compress]{article}

\usepackage{hyperref}
\usepackage{amsmath}
\usepackage{amsthm}
\usepackage{amsfonts}
\usepackage{amssymb}
\usepackage[all]{xy}
\usepackage{graphicx}
\usepackage{color}
\usepackage[mathscr]{eucal}
\usepackage[isu,bf]{caption}
\newlength{\xtrawidth}
\newlength{\xtraheight}
\numberwithin{equation}{section}

\def\e{\epsilon}

\def\clap#1{\hbox to 0pt{\hss#1\hss}}

{\end{list}}
{\everymath{\displaystyle\everymath{}}\array}%
{\endarray}
\newcommand{\lrstack}[3][t]{
  \ensuremath{
    \begin{array}[#1]{c}
      \multicolumn{1}{l}{\displaystyle{#2}\quad} \\[0.5em] 
      \multicolumn{1}{r}{\displaystyle\quad{#3}}
    \end{array}
  }}
\newcommand{\eqdef}{%
=}

\DeclareMathOperator{\Ext}{Ext}

\DeclareMathOperator{\rank}{rank}

\DeclareMathOperator{\Tr}{Tr}

\newcommand{\Z}{\mathbb{Z}}

\newcommand{\C}{\mathbb{C}}

\newcommand{\CP}[1]{\mathbb{P}^{#1}}

\newcommand{\Lsheaf}{\ensuremath{L}}
\newcommand{\Vsheaf}{\ensuremath{V}}

\newcommand{\dP}[1]{\ensuremath{\mathrm{dP}_{#1}}{}}
\newcommand{\ptset}{\ensuremath{\{\text{pt.}\}}}
\newcommand{\dual}{{\ensuremath\vee}}

\newcommand{\Osheaf}{\ensuremath{\mathscr{O}}}

\newcommand{\OB}[1]{\ensuremath{\Osheaf_{B_{#1}}}}
\newcommand{\OP}{\ensuremath{\Osheaf_{\CP1}}}
\newcommand{\Fsheaf}{\ensuremath{\mathscr{F}}}

\DeclareMathOperator{\FM}{FM}
\DeclareMathOperator{\supp}{supp}
\DeclareMathOperator{\Hsupp}{Hsupp}

\newcommand{\Rep}[1]{\ensuremath{\mathbf{\underline{#1}}}}
\newcommand{\barRep}[1]{\ensuremath{\overline{\Rep{#1}}}}
\newcommand{\detinv}{\operatorname{det}^{-1}}

\newcommand{\Ncal}{\ensuremath{\mathcal{N}}}

\begin{document}
\begin{titlepage}
  \begin{flushright}
    hep-th/0606241
    \\
    UPR 1158-T
  \end{flushright}
  \vspace*{\stretch{1}}
  \begin{center}
     \LARGE 
     Towards Realizing Dynamical SUSY Breaking in Heterotic Model Building
  \end{center}
  \vspace*{\stretch{2}}
  \begin{center}
    \begin{minipage}{\textwidth}
      \begin{center}
        \large         
        Volker Braun$^{1,2}$, 
        Evgeny I. Buchbinder$^{3}$,
        Burt A.~Ovrut$^{1}$
      \end{center}
    \end{minipage}
  \end{center}
  \vspace*{1mm}
  \begin{center}
    \begin{minipage}{\textwidth}
      \begin{center}
        ${}^1$ Department of Physics,
        ${}^2$ Department of Mathematics
        \\
        University of Pennsylvania,        
        Philadelphia, PA 19104--6395, USA
      \end{center}
      \begin{center}
        ${}^3$ School of Natural Sciences, 
        Institute for Advanced Study\\
        Einstein Drive, Princeton, NJ 08540
      \end{center}
    \end{minipage}
  \end{center}
  \vspace*{\stretch{1}}
  \begin{abstract}
    \normalsize 
    We study a new mechanism to dynamically break supersymmetry in the
    $E_8\times E_8$ heterotic string. As discussed recently in the
    literature, a long-lived, meta-stable non-supersymmetric vacuum
    can be achieved in an $\Ncal=1$ SQCD whose spectrum contains a
    sufficient number of light fundamental flavors. In this paper, we
    present, within the context of the hidden sector of the weakly and
    strongly coupled heterotic string, a slope-stable, holomorphic
    vector bundle on a Calabi-Yau threefold for which all matter
    fields are in the fundamental representation and are massive at
    generic points in moduli space. It is shown, however, that near
    certain subvarieties in the moduli space a sufficient number of
    light matter fields can occur, providing an explicit heterotic
    model realizing dynamical SUSY breaking. This is demonstrated for
    the low-energy gauge group $Spin(10)$. However, our methods
    immediately generalize to $Spin(N_{c})$, $SU(N_{c})$, and
    $Sp(N_{c})$, for a wide range of color index $N_{c}$.  Moduli
    stabilization in vacua with a positive cosmological constant is
    briefly discussed.
  \end{abstract}
  \vspace*{\stretch{5}}
  \begin{minipage}{\textwidth}
    \underline{\hspace{5cm}}
    \centering
    \\
    Email: 
    \texttt{vbraun, ovrut@physics.upenn.edu},
    \texttt{evgeny@sns.ias.edu}.
  \end{minipage}
\end{titlepage}

\tableofcontents

\section{Introduction}


Heterotic $M$-theory \cite{Horava:1995qa, Horava:1996ma,
  Witten:1996mz, Lukas:1997fg, Lukas:1998yy, Lukas:1998tt} provides a
promising framework to construct string theory vacua with the spectrum
of the supersymmetric standard model.  Recently, vacua of this kind
were obtained in heterotic compactifications on non-simply connected
Calabi-Yau manifolds~\cite{Braun:2005ux, Braun:2005bw, Braun:2005zv,
  Braun:2005nv, Braun:2006ae, Bouchard:2005ag}. One important task is
to understand how supersymmetry can be broken in these models. A
natural attempt would be to create a hidden sector with broken
supersymmetry and to communicate this breaking to the standard model
sector via some mediation mechanism.  Recently, a discussion of such
mechanisms in various string compactification and brane models was
presented in~\cite{Diaconescu:2005pc}. In~\cite{Intriligator:2006dd},
Intriligator, Seiberg, and Shih demonstrated that a class of $\Ncal=1$
SQCD theories generates dynamical SUSY breaking in a metastable
vacuum.  This class involves theories whose matter spectrum consists
solely of $N_f$ massive fundamental multiplets, where $N_{f}$ is in the
free magnetic range. The existence of this vacuum can then be
explicitly established by using the Seiberg dual
description~\cite{Seiberg:1994bz, Seiberg:1994pq,
  Intriligator:1995id}, in which the SUSY breaking vacuum appears at
weak coupling. In addition, these theories have $N_c$ supersymmetric
vacua so that the SUSY breaking vacuum is metastable.

It is important to understand whether this type of supersymmetry
breaking can be embedded in string theory, especially in realistic
compactifications and brane models with stable moduli.
In~\cite{Franco:2006es, Ooguri:2006pj} these questions were studied in
Type II string theory. In~\cite{Braun:2006em}, we began the study of
how dynamical SUSY breaking can be realized in realistic theories of
the $E_{8} \times E_{8}$ heterotic string. In this paper, we continue
this research, presenting all the requisite technical details and
proofs leading to the results in~\cite{Braun:2006em}. The obvious
approach is to construct vacua whose low-energy field theory satisfies
the criteria of~\cite{Intriligator:2006dd} in the hidden sector. Since
the desired low energy theory is non-chiral, we have to choose a
hidden sector vector bundle with vanishing third Chern class. The
spectrum of light particles is determined by the cohomology groups
with coefficients in different products of this vector bundle. As one
moves in the associated moduli space, some of the non-chiral matter
becomes massless on higher co-dimension subvarieties. Thus, the
first step would be to find a subvariety on which the massless
spectrum satisfies the representation and multiplicity criteria
of~\cite{Intriligator:2006dd}.  Then as we move slightly away from
this subvariety, the matter receives a small mass, which is the final
requirement in~\cite{Intriligator:2006dd}.  Unfortunately, moduli
spaces of Calabi-Yau manifolds and vector bundles are complicated and
it is usually difficult to prove the existence of subvarieties with
the requisite properties. In this paper, we will explicitly construct
one class of examples where this is achieved.  The structure group of
the vector bundle is chosen to be $SU(4)$, which leads to a low-energy
field theory with gauge group $Spin(10)$.  We show that, in this
example, it is possible to constrain the moduli in such a way that
$N_{f}$ fundamental multiplets of $SO(10)$, for any integer $N_{f}$,
obtain light masses whereas all other matter fields are heavy and can
be integrated out. This gives an example of a class of vacua of the
heterotic string whose low-energy field theories satisfying the
criteria of~\cite{Intriligator:2006dd}. It is important to note that
heterotic compactifications potentially have completely stabilized
moduli. We will not discuss this in the present paper.  Various
aspects of moduli stabilization in different heterotic models can be
found in~\cite{Buchbinder:2003pi, Gukov:2003cy, Buchbinder:2004im,
  Curio:2001qi, Becker:2004gw, Braun:2006th, Becker:2003yv,
  Becker:2003gq, Becker:2003sh, deCarlos:2005kh, Buchbinder:2005jy,
  Buchbinder:2006ab, Cardoso:2003af, LopesCardoso:2003sp,
  Blumenhagen:2005ga}.

This paper is organized as follows. In Section~\ref{sec:breaking}, we
review the criteria that theories with dynamical SUSY breaking must
satisfy~\cite{Intriligator:2006dd}. In Section~\ref{sec:embed}, a
general discussion of quadratic superpotentials for matter fields in
heterotic compactifications is presented.  The fact that the number of
light fields changes as we move in the moduli space means that there
exists a non-vanishing quadratic superpotential whose mass
coefficients are moduli dependent. The generic superpotential is one
which is cubic in the open string fields.  That is, it is a quadratic
function in the matter fields and a linear function in the vector
bundle moduli.  However, there can also be higher order contributions
to the superpotential of open string fields arising from integrating
out heavy Kaluza-Klein modes. Additionally, we give a brief discussion
of moduli stabilization and the possible relevance of the metastable
SUSY breaking for obtaining vacua with a small, positive cosmological
constant. In Section~\ref{sec:jumping}, it is shown how the matter
spectrum can change for special values of the moduli. This is the
basis for our choice of hidden sector vector bundle, which we describe
in Section~\ref{sec:model}.

To be as explicit as possible, we study the vector bundle and region
of moduli space leading to $N_{f}=8$ flavors. The Calabi-Yau threefold
is a double elliptic fibration over \dP9 (del Pezzo) surfaces, and the
$SU(4)$ vector bundle is constructed as a non-trivial extension with
building blocks pulled back from the two different \dP9 bases. The
slope-stability of the vector bundle is proven in
Subsection~\ref{sec:stab}, and follows from well-known results about
extensions of spectral cover bundles. This choice of the vector bundle
makes the study of cohomology groups tractable. On each \dP9 surface,
the cohomology groups of interest are localized at points.  The
dimensions of the cohomology groups becomes the number of points where
the supports of two such factors overlap. We also show that the
superpotential is cubic in the open string fields. By arranging the
supports in the right way, one can construct $\Ncal=1$, $SO(N_c)$ SQCD
with $N_{f}$ massive fundamentals satisfying the criteria
of~\cite{Intriligator:2006dd}. Hence, one can construct vector bundles
of this type where the number of light fundamental representations
lies in the appropriate range for dynamical symmetry breaking.

In Section~\ref{sec:explicit}, we present a mathematical proof of the
various details of the spectral cover used in Section~\ref{sec:model}.
In the conclusion, Section~\ref{sec:conclusion}, we discuss possible
extensions of our results. Finally, mathematical properties of the
support of line bundles and derived tensor products needed in our
analysis are presented in Appendices~\ref{sec:Lsupp}
and~\ref{sec:derived} respectively.


\section{Dynamical SUSY Breaking}
\label{sec:breaking}

In this section, we will give a brief review of dynamical
supersymmetry breaking following~\cite{Intriligator:2006dd}. We will
state certain necessary ingredients which will be used in later
sections.

The main example studied in~\cite{Intriligator:2006dd} was $\Ncal=1$,
$SU(N_c)$ SQCD with $N_f$ fundamental flavors $Q$, $\tilde{Q}$ in the
free magnetic range~\cite{Seiberg:1994bz, Seiberg:1994pq}
\begin{equation}
  N_c+1 \leq N_f < \frac{3}{2}N_c
  \,.
  \label{1.1}
\end{equation}
The flavors have a quadratic superpotential
\begin{equation}
  W= \Tr m M
  \,, 
  \label{1.2}
\end{equation}
where
\begin{equation}
  M = Q_f \cdot \tilde{Q}_g, \quad f, g=1, \dots, N_f
  \,,
  \label{1.25}
\end{equation}
so that they are all massive. This theory is known to have 
$N_c$ supersymmetric vacua with
\begin{equation}
  \left<M\right>=
  \big( \Lambda^{3N_c-N_f} \det m \big)^{1/N_c}
  m^{-1}
  \,,
  \label{1.3}
\end{equation}
where $\Lambda$ is the strong-coupling scale. In was shown
in~\cite{Intriligator:2006dd} that, in addition, this theory has a
metastable SUSY breaking vacuum. This was established by studying the
Seiberg dual~\cite{Seiberg:1994bz, Seiberg:1994pq} of the original
theory.  The Seiberg dual theory is $SU(N_f-N_c)$ SQCD with $N_f$
fundamentals $q$, $\tilde{q}$ and $N_f^2$ extra singlets $\Phi_f^g$. It
has a quadratic leading order K\"ahler potential and the superpotential
is given by (up to some field redefinition)
\begin{equation}
  W_\text{dual}=h \Tr q \Phi \tilde{q} -h \mu^2 \Tr \Phi
  \,
  \label{1.4}
\end{equation}
where $\mu=\sqrt{m\Lambda}$ and $h$ is a dimensionless parameter
defined in~\cite{Intriligator:2006dd}. For simplicity, we have assumed
that all eigenvalues of the mass matrix are equal.  This theory breaks
supersymmetry by a rank condition mechanism since F-flatness for
$\Phi$ requires that
\begin{equation}
  \tilde{q}^f q_g =\mu^2 \delta^f_g
  \,, 
  \label{1.5}
\end{equation}
which cannot be satisfied because the number of colors of the dual
theory $N_f-N_c$ is less than the number of flavors $N_f$. However, it
was shown in~\cite{Intriligator:2006dd} that there exists a metastable
SUSY breaking vacuum with
\begin{equation}
  V_{min}=N_c \big| h^2\mu^4 \big|
  \,, 
  \label{1.6}
\end{equation}
a result which can be trusted in the regime
\begin{equation}
  \epsilon \sim 
  \sqrt{ \left| \frac{m}{\Lambda} \right| } \ll 1
  \,.
  \label{1.7}
\end{equation}
Furthermore, as $\epsilon \rightarrow 0$ this state becomes very long-lived. For 
$\epsilon$ sufficiently small, the life-time of the meta-stable state can exceed
the age of the Universe, making these vacua of phenomenological interest.

These results were also generalized in~\cite{Intriligator:2006dd} for
SQCD with gauge groups $SO(N_c)$ and $Sp(N_c)$. In this paper, we will be 
particularly interested in $SO(N_c)$ theories. Hence we review
some important facts about them~\cite{Intriligator:2006dd,
  Intriligator:1995id}. $SO(N_c)$ SQCD has only one type of
fundamental representation $Q_f$. The tree-level superpotential is given by
eq.~\eqref{1.2} with
\begin{equation}
  M=Q_f \cdot Q_g
  \,.
  \label{1.75}
\end{equation}
The free magnetic range is defined by
\begin{equation}
  N_c-2 
  ~<~ 
  N_f 
  ~<~
  \frac{3}{2}(N_c-2)
  \,.
  \label{1.8}
\end{equation}
The Seiberg dual theory then has the (non-Abelian) gauge group
$SO(N_f-N_c+4)$ and the tree-level superpotential of the type
eq.~\eqref{1.4} with $\tilde{q}$ replaced by $q$. For $N_f=N_c-2$, the
Seiberg dual gauge group is $SO(2)\simeq U(1)$.  Thus, the dual theory
is really in the Coulomb phase. However, the SUSY breaking vacuum
still exists. Finally, there are special cases for $N_f=N_c-3$ and
$N_f=N_c-4$. A detailed investigation~\cite{Intriligator:2006dd}
reveals that they have SUSY breaking vacua as well.

To summarize, if the number of fundamentals is in the range
\begin{equation}
  \label{eq:Nfrange}
  N_c-4 
  ~\leq~ 
  N_f 
  ~<~
  \frac{3}{2}(N_c-2)
  \,,
\end{equation}
then the $SO(N_C)$ theory has a metastable SUSY breaking vacuum,
which can be trusted in the regime eq.~\eqref{1.7}. Details of dynamical SUSY
breaking in SQCD with the gauge group $Sp(N_c)$ can be found in
Subsection 6.3 of~\cite{Intriligator:2006dd}.


\section{Embedding in Heterotic Compactifications}
\label{sec:embed}

Compactifications of heterotic string theory or heterotic M-theory
provide a promising way of obtaining a realistic supersymmetric
standard model spectrum with stabilized moduli. The models
of~\cite{Intriligator:2006dd}, reviewed in the previous section, can
provide a mechanism to break supersymmetry in heterotic
compactifications.  Below we will give a general discussion of how
dynamical supersymmetry breaking can be embedded in heterotic
compactifications as the hidden sector.  In the next section, we will
present a concrete class of heterotic compactifications where the
spectrum satisfies the requisite properties
of~\cite{Intriligator:2006dd}.


\subsection{Quadratic Superpotentials for Matter Fields}

An important ingredient of dynamical SUSY breaking models is the tree
level quadratic superpotential. Therefore, it is important to discuss
how quadratic superpotentials for matter field can arise in heterotic
compactifications. Let $X$ be a compactification Calabi-Yau threefold
and $V$ be a vector bundle. The massless particle spectrum is
associated with the zero modes of the Dirac operator on $X$. Such zero
modes are in one-to-one correspondence with bundle-valued closed
differential $(0, 1)$-forms and, hence, bundle cohomology groups
$H^1(X, U)$, where $U$ can be $V$, $V^\dual$, $\wedge^2 V$, $\dots$.
Cohomology groups with coefficients in different $U$ bundles define
the massless states in the corresponding representations of the
low-energy gauge group in four dimensions.

However, the dimensions of these bundle cohomology groups are not a
topological invariant.  They depend on the location in the vector
bundle and complex structure moduli space.  As we move in these moduli
spaces, $h^{1}(X, U)$ can jump. This means that the corresponding
four-dimensional fields have a quadratic superpotential with the mass
depending on the vector bundle and complex structure moduli.
Somewhere in the moduli space these masses can vanish, thus increasing
the number of the massless particles.  If a compactification has some
chiral matter, then a certain number of fields will always stay
massless since they are protected by a topological invariant, the
Atiyah-Singer index.  On the other hand, the models reviewed in
Section~\ref{sec:breaking} are non-chiral. Hence, we are interested in
compactifications with no chiral matter. In this case, there are no
obvious obstructions to all matter multiplets having a non-vanishing
potential. One should expect, in compactifications with no chiral
matter, that every matter field will have a quadratic potential at a
generic point in moduli space. However, as we move in the moduli space
some fields can become light on higher co-dimension subvarieties.

Let us now discuss where quadratic potentials for matter fields can
come from.  Let $Q$ be a four-dimensional matter field transforming in
some representation $R$ of the low-energy gauge group, $\tilde{Q}$ be
a matter field in the conjugate representation $\bar{R}$ ($\tilde{Q}$
might coincide with $Q$ if $R$ is real) and $\phi$ represent vector
bundle moduli. All these fields correspond to $(0, 1)$-forms on $X$
with coefficients in the vector bundles $U_R, U_R^\dual$ and
ad($V$)respectively. Denote these forms as $\Psi_Q, \Psi_{\tilde{Q}}$
and $\Psi_{\phi}$. Upon dimensional reduction, these fields get a
cubic superpotential (see, for example,~\cite{Green:1987mn}) of the
form
\begin{equation}
  W=\lambda \phi \Tr Q \tilde{Q}
  \,.
  \label{2.1}
\end{equation}
The coefficients $\lambda$ depend on the complex structure and vector
bundle moduli and are given by
\begin{equation}
  \lambda=\int_{X} 
  \Omega \wedge 
  \Tr \Big(\Psi_{\phi} \wedge \Psi_{Q} \wedge \Psi_{\tilde{Q}} \Big)
  \,, 
\label{2.2}
\end{equation}
where $\Omega$ is the holomorphic $(3, 0)$-form.  If $Q$ and
$\tilde{Q}$ are Higgs fields, this superpotential represents a
$\mu$-term for them. Recently, such $\mu$-terms were computed in
realistic compactification scenarios in~\cite{Braun:2005xp,
  Braun:2006me, Bouchard:2006dn}.  The superpotential eq.~\eqref{2.1}
provides a generic mechanism for non-chiral matter to receive a mass
depending on various moduli. In addition, the open string fields can
also get a quartic superpotential of the form
\begin{equation}
  W \sim \phi \phi \Tr Q \tilde{Q}
  \,. 
  \label{2.3}
\end{equation}
Such a superpotential can arise after integrating out massive
Kaluza-Klein modes.  Indeed, let $\tilde{Q}_{KK}$ be a Kaluza-Klein
mode in the representation $\bar R$.  It can couple to $\phi$ and $Q$
through the a superpotential similar to eq.~\eqref{2.1}. In addition,
it has a quadratic superpotential with constant mass of order the
compactification scale.  Integrating $\tilde{Q}_{KK}$ out is
equivalent to eliminating auxiliary fields in supersymmetric field
theories. This procedure yields a quartic superpotential of the form
eq.~\eqref{2.3}.


\subsection{On Moduli Stabilization and Vacua with a Positive Cosmological Constant}

Eventually, we are interested in compactifications with stable moduli.
In this case, we can replace the complex structure and vector bundle
moduli with their vacuum expectation values (VEV), thus obtaining a
quadratic superpotential for the non-chiral matter. Let us consider a
compactification leading, at low energy, to a heterotic standard model
in the observable sector and to a hidden sector with gauge group
$SU(N)$, $SO(N)$, or $Sp(N)$. As an example, one can take the
structure group of the hidden sector vector bundle to be $SU(5)$, thus
obtaining another $SU(5)$ as the low energy gauge group. Another
example is to choose an $SU(4)$ structure group, leading to an
$SO(10)$ low energy gauge group in the hidden sector. We start our
analysis by ignoring all couplings to matter fields and finding a
supersymmetric AdS vacuum by solving
\begin{equation}
  D_\text{moduli}W_\text{moduli}=0
  \,. 
  \label{2.4}
\end{equation}
Questions concerning moduli stabilisation in heterotic
compactifications were studied in~\cite{Buchbinder:2003pi,
  Gukov:2003cy, Buchbinder:2004im, Curio:2000dw, Curio:2003ur,
  Curio:2001qi, Becker:2004gw, Braun:2006th, Becker:2003yv,
  Becker:2003gq, Becker:2003sh, deCarlos:2005kh, Buchbinder:2005jy,
  Buchbinder:2006ab}, and we will not review them in this paper.  For
our purposes, we assume that eq.~\eqref{2.4} stabilizes all the moduli
in a phenomenologically acceptable range. We further assume that the
moduli VEVs give the hidden sector fundamental matter, for all $N_f$
flavors in the free magnetic range, a small mass from the
superpotential eq.~\eqref{2.1}. All the remaining non-chiral matter
has very heavy mass and is integrated out. In the next section, we
will present an explicit example of a compactification with such
properties.  By the results of~\cite{Intriligator:2006dd}, the
supersymmetry will then be broken dynamically in the hidden sector.
This supersymmetry breaking is communicated to the standard model
sector by one of the mediation mechanisms (see, for
example,~\cite{Giudice:1998bp} for a review). The metastable SUSY
breaking vacuum obtained in~\cite{Intriligator:2006dd} can be trusted
in the regime where $\epsilon \ll 1$ and where one can neglect the
$\frac{1}{M_{Pl}}$ contributions to the potential energy. This
dynamical SUSY breaking also has an obvious effect on the cosmological
constant. Let $W_0$ be the value of the moduli superpotential in the
solution eq.~\eqref{2.4}. It produces a negative contribution to the
cosmological constant of order $-3\frac{|W_0|^2}{M^2_{Pl}}$. On the
other hand, the matter in the hidden sector in the metastable SUSY
breaking vacuum gives a positive correction to the cosmological
constant.  Depending on the relative values of $m$ and $W_0$, one can
obtain a non-supersymmetric vacuum with a negative, vanishing, or
positive cosmological constant. In particular, vacua with a small,
positive cosmological constant can potentially be obtained this way.
This important physics is model dependent and goes beyond the range of
this paper. Hence, we will not discuss it here but leave it for future
research.


\section{Mass Terms and Discontinuous Cohomology}
\label{sec:jumping}

Before we are going to delve into the technicalities of our model, let
us first describe the underlying idea of the construction. As
described in Section~\ref{sec:embed}, we want a hidden sector that
contains no massless matter fields at a generic point in the moduli
space, but does contain massless matter for special values of the
moduli. This is possible since the sheaf cohomology that computes the
spectrum is not a topological invariant, but can in fact change as one
varies the vector bundle moduli~\cite{Donagi:2004qk, Donagi:2004ia,
  Donagi:2004su, Donagi:2004ub}. The simplest such ``jump'' occurs
already for an elliptic curve.

Let us start by reviewing this case, and take 
\begin{equation}
  E = \C \big/ \big( \Z + \tau \Z \big)
\end{equation}
to be an elliptic curve, and let us fix the point $0+0i=o\in E$. An
elliptic curve with origin is, in fact, a group: The group law
$\boxplus$ on the points of $E$ is addition in $\C$ modulo the lattice
$\Z + \tau \Z$. Now the divisors on $E$ are formal $\Z$-linear
combinations of points, and every line bundle can be written as
\begin{equation}
  \Osheaf_E\Big({\textstyle \sum_{i=1}^n p_i - \sum_{j=1}^m q_j }\Big)
  \,,\qquad p_i, q_j \in E
\end{equation}
But not all such line bundles are distinct, and the isomorphism
classes of holomorphic line bundles on $E$ can be labeled by the two
numbers
\begin{subequations}
\begin{align}
  n-m ~&\in~ \Z
  \\
  \big( \boxplus_{i=1}^n p_i \big)
  \boxminus
  \big( \boxplus_{j=1}^m q_j \big)
  ~&\in~ E
\end{align}
\end{subequations}
Depending on these two invariants, there are four cases to
distinguish. They are
\begin{equation}
  \renewcommand{\arraystretch}{1.3}
  \begin{tabular}{c|cccc}
    & Case 1 & Case 2 & Case 3 & Case 4
    \\ \hline
    $n-m$
    & $> 0$ & $=0$ & $=0$ & $< 0$
    \\
    $\big( \boxplus_{i=1}^n p_i \big)
    \boxminus
    \big( \boxplus_{j=1}^m q_j \big)$
    & any & $\in E-\{o\}$ & $= o$ & any
    \\ 
    $\dim H^0\big( \sum_{i=1}^n p_i - \sum_{j=1}^m q_j \big)$
    & $n-m$ & $0$ & $1$ & $0$
    \\
    $\dim H^1\big( \sum_{i=1}^n p_i - \sum_{j=1}^m q_j \big)$
    & $0$ & $0$ & $1$ & $m-n$
    \,.
  \end{tabular}
\end{equation}
In particular, we are interested in the $n-m=0$ case. Then the line
bundle has vanishing first Chern class, but there are still two
possibilities. Either the line bundle is the trivial line bundle
$\Osheaf_E= E \times\C$, or the line bundle is of the form
$\Osheaf_E(p-o)$ for some $p\not=o\in E$. In the first case
$H^0(E,\Osheaf_E)=H^1(E,\Osheaf_E)=\C$, while in the latter case all
cohomology groups vanish.

The underlying idea of the spectral cover construction is to apply
this fiberwise to an elliptic fibration. Consider a spectral curve $C$
that is a $k$-fold cover of the base, and let $\sigma$ be the zero
section of the elliptic fibration. Then $C$ intersects a generic fiber
$f$ in $k$ separate points $C_1$, $\dots$, $C_k$, and $\sigma$
intersects the fiber $f$ in the single point $o\in F$. The
Fourier-Mukai transform constructs a rank $k$ vector bundle whose
restriction to $f$ is
\begin{equation}
  \label{eq:FMC}
  \FM( \Osheaf_C )\big|_f = 
  \Osheaf_f(C_1-o) 
  \oplus \cdots \oplus
  \Osheaf_f(C_k-o) 
  \,.
\end{equation}
Obviously, the cohomology of $\FM( \Osheaf_C )\big|_f$ vanishes unless
one of the points $C_1$, $\dots$, $\C_k$ coincides with $o$. But
according to the Leray spectral sequence (see, for example,
~\cite{MR507725}), the cohomology groups of $\FM( \Osheaf_C )$ can be
computed in terms of the fiberwise cohomology. If the latter vanishes,
then the cohomology of $\FM( \Osheaf_C )$ has to vanish as well.

Note that it is not enough if only the cohomology at generic fibers
vanishes, but only if it vanishes at every fiber. Since the
intersection points $C\cap f$ vary as we vary the fiber $f$, we expect
that there are some fibers where $C_1=o$ or $C_2=o$ or $\dots$ or
$C_k=o$. In terms of the zero section $\sigma$ of the elliptic fibration,
these points are $C\cdot \sigma$. As $C_i=o$ is one complex equation, these
special fibers occur in codimension one on the base. If we were to
consider an elliptically fibered Calabi-Yau threefold, then the
complex $2$-dimensional base will in general contain a curve which
supports cohomology groups. Instead, we will take the Calabi-Yau
threefold $X$ to be fibered over $\CP1$,
\begin{equation}
  X 
  \stackrel{pr}{\longrightarrow}
  \CP1
  \,,
\end{equation}
such that a generic fiber
\begin{equation}
  \label{eq:pr}
  pr^{-1}\big(\ptset\big) \simeq E_1 \times E_2
\end{equation}
factors into the product of two elliptic curves. Then we arrange
spectral cover-like bundles on $E_1$ and $E_2$ separately, that is,
construct a bundle such that the restriction to
$pr^{-1}\big(\ptset\big)$ is
\begin{multline}
  \label{eq:absurfbundle}
  \Big( 
  \Osheaf_{E_1}(C_1-o) 
  \oplus \cdots \oplus
  \Osheaf_{E_1}(C_k-o) 
  \Big) 
  \boxtimes 
  \Big( 
  \Osheaf_{E_2}(D_1-o) 
  \oplus \cdots \oplus
  \Osheaf_{E_2}(D_l-o) 
  \Big) 
  \,,\\
  C_1,\dots,C_k \in E_1
  \,,\quad
  D_1,\dots,D_l \in E_2
  \,.
\end{multline}
Generically none of the points $\{C_1,\dots,C_k\}$ and none of the
points $\{D_1,\dots,D_l\}$ coincides with $o$, and the cohomology
along the fiber direction vanishes. Only if $C_i=o=D_j$ simultaneously
for some $i=1,\dots,k$, $j=1,\dots,l$ then the cohomology of the
bundle eq.~\eqref{eq:absurfbundle} is non-vanishing. But that yields
two complex equations on the $1$-dimensional base $\CP1$, which has no
solutions in general. Only specially designed bundles then have
non-zero cohomology groups, while any small deformation will lead to
vanishing cohomology.


\section{The Compactification}
\label{sec:model}

\subsection{The Calabi-Yau Threefold}
\label{sec:CY}

In this section, we will present a concrete model of the hidden sector
satisfying the criteria for dynamical SUSY breaking. Since we are only
interested in the supersymmetry breaking in the hidden sector, we will
not specify the visible sector and the five-brane structure. In our
model, we choose the Calabi-Yau threefold $X$ to be a double elliptic
fibration~\cite{SchoenCY, Donagi:2000zs, Ovrut:2002jk, Donagi:2003tb,
  Braun:2004xv}
\begin{equation}
  X=B_1 \times_{\CP1} B_2, 
  \label{3.1}
\end{equation}
where
\begin{equation}
  B_1 \simeq \dP9
  \,, \quad
  B_2 \simeq \dP9
  \label{3.1.1}
\end{equation}
are two rational elliptic ($\dP9$) surfaces. We will denote
projections by $\pi_i =X \to B_i$ and $\beta_i =B_i \to \CP1$, $i=1,
2$, yielding a commutative square
\begin{equation}
  \label{eq:projections}
  \vcenter{\xymatrix@!0@=12mm{
      \dim_\C=3: && & X \ar[dr]^{\pi_2} \ar[dl]_{\pi_1} \\
      \dim_\C=2: && B_1 \ar[dr]_{\beta_1} & & 
        B_2 \ar[dl]^{\beta_2} \\
      \dim_\C=1: && & \CP1
      \,.
  }}
\end{equation}
The fibers of these projections are generically elliptic curves, with
some degenerate fibers. The Abelian surface fibration of
Section~\ref{sec:jumping}, eq.~\eqref{eq:pr} is simply
$pr=\beta_1\circ\pi_1 = \beta_2\circ\pi_2$. Let us state some
properties of the homology group of curves $H_2 (B_i, \Z)$ which we
will be using.  A \dP9 surface is obtained by blowing up nine points
of $\CP2$.  From $\CP2$ we inherit the class of the hyperplane divisor
$\ell$, and each blow-up adds one exceptional divisor.
Hence\footnote{The construction involves two distinct \dP9 surfaces
  $B_1$ and $B_2$. Hence, strictly speaking, one needs to distinguish
  their divisors by labeling them differently. However, it will always
  be clear from the context which surface we are referring to.
  Therefore, we suppress this extra label.},
\begin{equation}
  H_2 \big(B_i, \Z\big) = 1 + 9 = 10
  \,.
\end{equation}
We denote these 9 exceptional divisors $e_i$, $i=1, \dots, 9$. The
intersection numbers of these classes are
\begin{equation}
  \ell \cdot \ell =1
  \,, \quad 
  e_i \cdot e_j = -\delta_{ij}
  \,, \quad 
  e_i \cdot \ell =0
  \,.
  \label{3.1.2}
\end{equation}
Obviously the determinant of the intersection matrix is $-1$, and
therefore the classes $\ell,\e_1,\dots,e_9$ are an integral basis for
the homology lattice. In this basis the fiber class of the \dP9
elliptic fibration reads
\begin{equation}
  f = 3 \ell -\sum_{i=1}^9 e_i
  \,. 
  \label{3.1.3}
\end{equation}
Each exceptional divisor $e_i$ is a section of \dP9 since it
intersects the fiber $f$ at one point.  We will choose $e_9$ to be the
zero section.

Finally, we need the even cohomology ring to compute Chern classes. It
is generated by the pull-backs
\begin{equation}
  \label{eq:HevGen}
  \begin{aligned}
    \lambda^1 
    \eqdef&~ 
    \pi_1^\ast(\ell)
    \,, \qquad & 
    \epsilon_i^1 
    \eqdef&~ 
    \pi_1^\ast(e_i)
    \,,\quad 
    i=1,\dots,9
    \,, \\
    \lambda^2
    \eqdef&~ 
    \pi_2^\ast(\ell)
    \,, & 
    \epsilon_i^2 
    \eqdef&~ 
    \pi_2^\ast(e_i)
    \,,\quad 
    i=1,\dots,9
    \,,
  \end{aligned}
\end{equation}
see~\cite{Donagi:2000zs, Ovrut:2003zj, Braun:2005zv}. The $T^4$ fiber
can be expressed in two different ways, yielding the relation
\begin{equation}
  \label{eq:HevRel1}
  3 \lambda^1 -\sum_{i=1}^9 \epsilon_i^1
  ~=~
  3 \lambda^2 -\sum_{i=1}^9 \epsilon_i^2
  \,.
\end{equation}
In addition, there are quadratic relations that are inherited from the
base \dP9 surfaces
\begin{equation}
  \label{eq:HevRel2}  
  \begin{aligned}
    \big( \lambda^1 \big)^2 =&~ - \big( \epsilon_1^1 \big)^2 
    \,, \qquad & 
    \big( \lambda^2 \big)^2 =&~ - \big( \epsilon_1^2 \big)^2 
    \,, \\    
    \lambda^1 e_i^1 =&~ 0
    \,, \qquad & 
    \lambda^2 \epsilon_i^2 =&~ 0    
    \,,&\quad i=&~1,\dots, 9
    \,, \\
    \epsilon_i^1 \epsilon_j^1 =&~ 
    \delta_{ij}
    \big( \epsilon_1^1 \big)^2 
    \,, \qquad & 
    \epsilon_i^2 \epsilon_j^2 =&~ 
    \delta_{ij}
    \big( \epsilon_1^2 \big)^2 
    \,,&\quad i,j=&~1,\dots, 9
    \,, \\
  \end{aligned}
\end{equation}
and one set of relations that involves both \dP9 surfaces,
\begin{equation}
  \label{eq:HevRel3}
  \big( 
  \epsilon_i^1 - \epsilon_j^1
  \big) 
  \big( 
  \epsilon_k^2 - \epsilon_l^2
  \big) 
  = 0
  \,,\quad
  i,j,k,l=1,\dots,9
  \,.
\end{equation}
To summarize, the even cohomology groups are 
\begin{equation}
  \label{eq:Hev}
  H^\text{ev}\big(X,\Z\big) = 
  \Z[ 
  \lambda^1,
  \lambda^2,
  \epsilon_1^1,\dots,\epsilon_9^1,
  \epsilon_1^2,\dots,\epsilon_9^2
  ] \Big/
  \big\{
  \text{Relations eqns.~}
  \eqref{eq:HevRel1}, 
  \eqref{eq:HevRel2}, 
  \eqref{eq:HevRel3}
  \big\}
  \,.
\end{equation}

\subsection{The Vector Bundle}
\label{sec:bundle}

Having described the base space $X$, we now construct a slope-stable,
holomorphic vector bundle $V$ with structure group $SU(4)$ and
vanishing third Chern class. Turning on such an instanton in the
hidden sector $E_8$ gauge group breaks it to $Spin(10)$. There are two
types of matter fields that appear in four dimensions, one can have
multiplets transforming as $\Rep{16}$, $\barRep{16}$, and $\Rep{10}$
of $Spin(10)$. Their number is given by $h^1(X, V)$, $h^1(X,
V^\dual)$, and $h^1(X, \wedge^2 V)$, respectively. Since we chose the
third Chern class of $V$ to be zero, the number of $\Rep{16}$ and the
number of $\barRep{16}$ is the same.

Let us now describe the vector bundle $V$.  We construct the rank $4$
vector bundle $V$ as a non-trivial extension of a line bundle and a
rank $3$ vector bundle, that is, 
\begin{equation}
  \label{eq:Vext}
  0 
  \longrightarrow
  V_1 
  \longrightarrow
  V
  \longrightarrow
  V_3 
  \longrightarrow
  0
  \,.
\end{equation}
The rank $3$ bundle $V_3$ will be
\begin{equation}
  \label{eq:V3def}
  V_3 =
  \pi_1^* \big(L\big) \,\otimes\, \pi_2^* \big(W\big)
  \,, 
\end{equation}
where $L$ is a line bundle on $B_1$ and $W$ is a rank $3$ vector
bundle on $B_2$ defined as follows. The line bundle is
\begin{equation}
  \label{eq:Ldef}
  L=\OB1(e_1-e_9)
  \,.
\end{equation}
Really we could use the difference of any two sections that do not
intersect, but for definiteness we will use the exceptional divisors
$e_1$ and $e_9$.

Furthermore, we define the rank three vector bundle $W$ using the
spectral cover construction. The spectral curve $C_W$, see~\cite{Friedman:1997yq},
is taken to be an irreducible curve in the linear system
\begin{equation}
  \label{eq:CWdef}
  C_{W} \in \Gamma \OB2\big( \ell +f \big)
  \,.
\end{equation}
From eqns.~\eqref{3.1.2} and~\eqref{3.1.3} it follows that $\ell$
intersects $f$ at three points and, thus, is a triple cover of the
base $\CP1$. In addition to the spectral curve $C_W$, we also have to
specify a line bundle $N_W$ on $C_W$. For simplicity, we take $N_W$
to be the trivial line bundle on $C_W$
\begin{equation}
  N_W = \Osheaf_C
  \,. 
  \label{3.6}
\end{equation}
The stable rank $3$ vector bundle $W$ is then obtained by the
Fourier-Mukai transform of $(C_W, N_W)$~\cite{Friedman:1997yq,
  DonagiPrincipal},
\begin{equation}
  \label{eq:Wdef}
  W \eqdef
  \FM_{B_2}\big( \Osheaf_C \big)
  \,.
\end{equation}
Using the action of the Fourier-Mukai transform on the level of Chern
classes which were worked out in~\cite{Donagi:2000fw}, we find
\begin{equation}
  \rank(W) =3
  \,,\quad
  c_1(W) = 
  \ell -3e_9 -8f
  \,,\quad
  c_2(W) = 0;
  \,.
\end{equation}
Note that $W$ is a $U(3)$ vector bundle with non-trivial $U(1)$ part
\begin{equation}
  \det W=\OB2(\ell -3e_9 -8f) 
\end{equation}
In particular, $\wedge^2 W$ is not isomorphic to $W^\dual$.

Using again techniques developed in~\cite{Donagi:2000fw}, one finds that
the spectral cover of $\wedge^2 W$ is in the linear system
\begin{equation}
  C_{\wedge^2 W} \in \Gamma \OB2\big( -2 \ell +9 e_9 +14 f \big)
  \,.
  \label{3.7}
\end{equation}
To make sure that $V$ has structure group $SU(4)$, we finally pick the
line bundle $V_1$ to be
\begin{equation}
  \label{eq:V1def}
  V_1 = 
  \pi_1^* \big( L^{-3} \big) \otimes 
  \pi_2^* \big( \detinv W \big)
  \,. 
\end{equation}
This choice of $V_1$ guarantees that the first Chern class of $V$ vanishes.

\subsection{Chern Classes and Stability}
\label{sec:stab}

Knowing the even cohomology ring explicitly eq.~\eqref{eq:Hev}, one
can easily compute all relevant Chern classes. One finds
\begin{equation}
  \rank(V) = 4
  \,, \quad
  c_1(V) = 0 
  \,, \quad
  c_2(V) = 12 \big(\lambda^1\big)^2 + 8 \big(\lambda^2\big)^2
  \,, \quad
  c_3(V) = 0
  \,.
\end{equation}
Therefore, the gauge and gravity contribution to the heterotic anomaly
equation for some visible sector bundle $V_\text{vis}$ reads
\begin{multline}
  c_2\big(TX\big) 
  - c_2\big(V_\text{vis}\big) 
  - c_2\big(V\big) 
  = \\ =
  \Big(12 \big(\lambda^1\big)^2 + 12 \big(\lambda^2\big)^2 \Big)
  - 
  \Big( 12 \big(\lambda^1\big)^2 + 8 \big(\lambda^2\big)^2 \Big)
  - c_2\big(V_\text{vis}\big) 
  = \\ =
  4 \big(\lambda^2\big)^2
  - c_2\big(V_\text{vis}\big)   
  \,,
\end{multline}
where $\big(\lambda^i\big)^2$ is the fiber class of $\pi_i$, $i=1,2$,
and hence an effective curve. We conclude that, depending on
$V_\text{vis}$, it is possible to cancel the heterotic anomaly without
introducing anti-five-branes. 

To show that $V$ is slope-stable for some suitable K\"ahler class, we
only have to satisfy~\cite{Friedman:1997ih, Donagi:2000zs}
\begin{enumerate}
\item The extension eq.~\eqref{eq:Vext} is not split.
\item The slope of $V_1$ is negative.
\end{enumerate}
We are going to compute the extensions in
Subsection~\ref{sec:cohomology}, and find that there are non-trivial
extensions. Finally, if one prefers to work in a region of the
K\"ahler moduli space where the slope $\mu(V_1)$ of $V_1$ is positive,
then
\begin{equation}
  \mu\big(V_1\big) > 0 
  \quad \Leftrightarrow \quad
  \mu\big(V_3\big) < 0   
  \,.
\end{equation}
In that case, one can just reverse the extension eq.~\eqref{eq:Vext}.
It turns out that for our bundle $V$ this does not influence any
cohomology groups. Hence, in one way or the other $V$ is slope-stable.


\subsection{Localization of Cohomology}

In this subsection, we will review some basics of sheaf cohomology and
how it applies to the vector bundles $V$ and $\wedge^2 V$ which we are
using throughout this paper. A detailed consideration of them in a
similar geometry can be found, for example, in subsections 7.3, 7.4
of~\cite{Donagi:2004ia}. Let 
\begin{equation}
  X \stackrel{\pi}{\longrightarrow} B
\end{equation}
be an elliptically fibered manifold, and $U$ be a slope-stable,
holomorphic vector bundle obtained via the spectral cover
construction~\cite{Friedman:1997yq, DonagiPrincipal}. In particular,
we assume that the restriction $U|_F$ to a generic fiber is regular
semistable and of degree $0$.  Applying the Leray spectral sequence to
that case (see, for example,~\cite{MR507725}), one finds that the
cohomology groups with coefficients in $U$ are localized in a
codimension two subvariety of $X$. As we saw in
Section~\ref{sec:jumping}, the localized cohomology groups correspond
to the intersection points
\begin{equation}
  C_U \cdot \sigma
  \,, 
  \label{3.9}
\end{equation}
where $C_U$ is the spectral cover\footnote{The result for $\wedge^2 U$
  is identical with $C_U$ being replaced by the spectral cover
  $C_{\wedge^2 W}$ of $\wedge^2 W$.} and $\sigma$ is the zero section.
In our case the Leray spectral sequence determines the cohomology
groups of $U$ on $X$ in terms of the cohomology of certain torsion
sheaves on $B$ with support on $\pi\big( C_U \cdot \sigma\big)$. This
torsion sheaf happens to be
\begin{equation}
  R^1 \pi_\ast U
  \,,
  \label{3.10}
\end{equation}
the sheaf whose ``fiber'' over a point $p\in B$ is $H^1\big(f_p,
U|_{f_{p}}\big)$, where $f_p=\pi^{-1}(p)$ is the fiber at the point
$p$. 
Hence, 
\begin{equation}
  H^1\big( X, U\big) = H^0\big( B, R^1\pi_\ast U\big)
  \,.
\end{equation}
As we discussed in Section~\ref{sec:jumping}, the ``fiber'' dimension
of $R^1\pi_\ast U$ is generically zero but can jump occasionally,
hence the $R^1\pi_\ast U$ is only a coherent sheaf and not a vector
bundle.

In our case, the building blocks of the bundle $V$ are vector bundles
on the \dP9 surfaces. It the following, it will be useful to
specialize the above to the case where the total space is the surface
$B_i$ with projection $\beta_i:B_i\to\CP1$. Let us denote the spectral
curve $C_{U_i}$ and the corresponding bundle $U_i$. In that case,
$C_{U_i} \cdot \sigma$ consists of a certain number of points. The
sheaf $R^1 \beta_{i*} U$ is the skyscraper sheaf supported at these
points and zero everywhere else. At each of these points $R^1
\beta_{i\ast} U_i$ is just $\C$. Thus\footnote{Here we are assuming
  for simplicity that the zero section $\sigma$ does not meet any
  singularities of the spectral curve $C_U$. If they do meet in
  singular points it is still true that the cohomology is supported at
  these points, one just has to be careful with the multiplicities.},
\begin{equation}
  H^1\big( B_i, U_i \big)
  =
  H^0\big( \CP1, R^1 \beta_{i\ast} U_i \big)
  =
  H^0\big( C_U \cdot \sigma, \C \big)
  \label{3.11}
\end{equation}
is nothing else than the number of points where the spectral cover
intersects the zero section. All higher cohomology vanish, and the
details of $R^1 \beta_{i\ast} U_i$ become irrelevant. Let us denote by
$\Hsupp(U_i)$ the points where the cohomology of $U_i$ is supported,
that is,
\begin{equation}
  \Hsupp(U_i) = \supp\big( R^1\beta_{i\ast} U_i \big)
  = \beta_i\big( C_{U_i} \cap \sigma \big)
  ~\subset\CP1
  \,.
\end{equation}
In the next section, we will be interested in a special limit where
points in $\Hsupp(U_i)$ collide. For that case one has to count the
points with multiplicities. Finally, we remark that
\begin{equation}
  \Hsupp(U_i^\dual) =
  \beta_i\Big( (\boxminus C_{U_i}) \cap \sigma \Big) =
  \beta_i\big( C_{U_i} \cap \sigma \big) =
  \Hsupp(U_i)  
  \,.
\end{equation}

In this paper, we will often encounter the case where the bundle on
the threefold $X=B_1\times_{\CP1}B_2$ is the tensor product of bundles
pulled back from $B_1$ and $B_2$, that is,
\begin{equation}
  \label{eq:U1U2tensorcase}
  U =
  \pi_1^\ast \big(U_1\big) \otimes 
  \pi_2^\ast \big(U_2\big)
\end{equation}
Such a vector bundle is, when restricted to a $T^4$-fiber of the
fibration $pr=\beta_1\circ\pi_1=\beta_2\circ\pi_2$, of the form
eq.~\eqref{eq:absurfbundle}. Hence the discussion at the end of
Section~\ref{sec:jumping} applies, and we expect $U$ to have no
cohomology for generic values of the moduli. Its cohomology can be
computed by applying the Leray spectral sequence twice, pushing down
from $X$ via $B_i$ to $\CP1$. One obtains\footnote{For technical
  reasons, we compute $H^2(X,U)$. This is explained in
  Appendix~\ref{sec:derived}.}
\begin{equation}
  \label{eq:Ucoh}
  \begin{split}
  H^2\big( X,U \big)
  ~=&~
  \left\{
  \begin{array}{c}
    H^1\Big( 
    U_1 \otimes
    \beta_1^\ast \circ R^1 \beta_{2\ast} \big(U_2\big) 
    \Big)
    \\
    H^1\Big( 
    \beta_2^\ast \circ R^1 \beta_{1\ast} \big(U_1\big) 
    \otimes U_2 \Big)
  \end{array}
  \right\}
  = \\ =&~
  H^0\Big( 
  R^1 \beta_{1\ast} \big(U_1\big) 
  \otimes 
  R^1 \beta_{2\ast} \big(U_2\big) 
  \Big)
  \end{split}
\end{equation}
where we either push down via $B_1$ or $B_2$.  Obviously, the
cohomology of $U$ is supported at the intersection $\Hsupp(U_1) \cap
\Hsupp(U_2)$. If $\Hsupp(U_1)$ is distinct from $\Hsupp(U_2)$, we
immediately get
\begin{equation}
  R^1 \beta_{1\ast} \big(U_1\big) 
  \otimes 
  R^1 \beta_{2\ast} \big(U_2\big)  
  = 0
  \quad \Rightarrow\quad
  H^2\big(X,U\big) = 0
  \,.
  \label{3.12}
\end{equation}
In general, $h^1 (X,U)$ is given by the number of points common to
both supports, that is (counted with appropriate multiplicities),
\begin{equation}
  \dim H^2\big(X,U\big) = 
  \big| \Hsupp(U_1)\cap\Hsupp(U_2) \big|
  \,.
\end{equation}
Finally, take the index
\begin{equation}
  \chi(U) = \sum_{i=0}^3 H^i\big(X,U\big)
\end{equation}
to be zero. This is always the case here, since we construct bundles
whose cohomology groups vanish at a generic point in the moduli space.
The index is unchanged as one changes the moduli, so if $H^2(X,U)$
jumps then $H^1(X,U)$ has to jump as well to compensate,
\begin{equation}
  \label{eq:Hsupp}
  H^1\big(X,U\big) \simeq H^2\big(X,U\big)
  = \big| \Hsupp(U_1)\cap\Hsupp(U_2) \big|
  \,.
\end{equation}
Here we used that $H^0(X,U)=0=H^3(X,U)$ as required by stability.

Let us now apply these results to calculating cohomology $H^1(V)$. In
order to reproduce the theory reviewed in Section~\ref{sec:breaking},
we should get $H^1(V)=0$. From the long exact sequence of cohomology
associated with the sequence eq.~\eqref{eq:Vext}, we find that
$H^1(V)=0$ if $H^*(V_1)=H^*(V_3)=0$. Let us show that this is indeed
the case at a generic point in the moduli space. Let us start with
$H^*(V_3)$.  Since the definition of $V_3$ in eq.~\eqref{eq:V3def} is
of the form eq.~\eqref{eq:U1U2tensorcase}, we can simply apply the
previous discussion. One finds that the support of $H^*(V_3)$ is
\begin{equation}
  \Hsupp(L) \cap \Hsupp(W)
  \,. 
  \label{3.13}
\end{equation}
Using eqns.~\eqref{3.1.2},~\eqref{3.1.3} and~\eqref{eq:CWdef}, we find
that the support of the cohomology of $W$, $\Hsupp(W)$, is given by
\begin{equation}
  C_W \cdot e_9 = 1
  \label{3.14}
\end{equation}
The precise location of this point depends on the moduli of $W$. To
obtain $\Hsupp(L)$ we have to calculate the sheaf $R^1\beta_{1*}L$.
This is performed in Appendix~\ref{sec:Lsupp}, and the result is
\begin{equation}
  R^1\beta_{1*} L = 0 
  \quad \Rightarrow \quad 
  \Hsupp(L) = \emptyset
\end{equation}
Thus, $H^*(V_3)$=0. The cohomology of $V_1$ is supported at
\begin{equation}
  \Hsupp\big(L^3\big) \,\cap\, \Hsupp\big(\det W\big). 
  \label{3.15}
\end{equation}
One can show that neither support in eq.~\eqref{3.15} is empty. However,
for generic choice of the complex structure of \dP9 and the bundle
moduli of $W$ the intersection is empty. Thus, at a generic point in
the moduli space $H^*(V_1)$ vanishes and so does $H^*(V)$.  This is
not surprising. As discussed before, in models with no chiral matter
one can expect that all matter has a quadratic superpotential. Thus we
can achieve that all particles transforming as $\Rep{16}$ and
$\barRep{16}$ receive a large mass and are integrated out.


\subsection{Extensions and the Spectrum of Fundamentals}
\label{sec:cohomology}

We want $V$ to be a non-trivial extension. For this we need
\begin{equation}
  \label{3.16}
  \Ext^1 \big(V_3, V_1 \big) \neq 0
  \,. 
\end{equation}
This is equivalent to 
\begin{equation}
  \label{3.17}
  H^1 \big(X, V_1 \otimes V_3^\dual \big) = 
  H^1 \Big(X, 
  \pi_1^\ast\big( L^{-4} \big) \otimes 
  \pi_2^\ast\big( W^\dual \otimes \detinv W \big)  
  \Big) 
  \neq 0
  \,. 
\end{equation}
To apply the discussion in the previous subsection we have to
understand the intersection of $\Hsupp(L^4)$ and $\Hsupp(W^\dual
\otimes \detinv W)$.  Despite the fact that the cohomology of $L$ has
vanishing support, the line bundle $L^2$ has non-trivial cohomology.
In Appendix~\ref{sec:Lsupp}, it is shown that $\Hsupp(L^2)$ consists
of three points on $\CP1$. Let us denote them by $q_1, q_2, q_3$. That
is,
\begin{equation}
  \Hsupp\big(L^2\big)=\big\{
  q_1, q_2, q_3
  \big\}
  \,.
  \label{3.18}
\end{equation}
The actual location of these points depends on the complex 
structure of $X$.
Furthermore, $\Hsupp\big(L^4\big)$ contains fifteen points. It can be
shown that these fifteen points contain $q_1, q_2, q_3$ each with
multiplicity one plus $12$ other points whose location is completely
generic. Let us denote these points by 
\begin{equation}
  \Hsupp\big(L^4\big)=\big\{
  q_1, q_2, q_3, s_1, s_2, \dots, s_{12} 
  \big\}
  \,.
\label{3.18.5}
\end{equation}
Later in this section we will need to know the cohomology support of
the bundle 
\begin{equation}
  \wedge^2 W = W^\dual \otimes \det W
  \,.
\end{equation}
According to our discussion in the previous subsection, it is given by
$C_{\wedge^2 W} \cdot e_9$.  Using eqns.~\eqref{3.1.2}, \eqref{3.1.3},
and~\eqref{3.7} these curves intersect in $5$ points. 
The location of these points depends on the moduli of $W$ 
and the complex structure of $X$.
In the next
section we will explicitly demonstrate that there exist a regime in
the moduli space where two points appear with multiplicity two. For
purposes that will be clear later on, we want to work in this case
where
\begin{equation}
  \Hsupp\big(\wedge^2 W\big) =\big\{2p_1, 2p_2, p_3\big\}
  \label{3.19}
\end{equation}
for some points $p_1, p_2, p_3\in \CP1$. 
Later, in this subsection, we will see that this choice leads to the 
spectrum with the number of fundamentals $N_f$ equal to eight. We found that
it is the most instructive to do this case in detail. In the next subsection,
we will discuss how different numbers of flavors can be obtained. In fact, 
some of them will be found as a simple modification of the $N_f=8$ case.
Finally, consider $W^\dual
\otimes \detinv W = (W\otimes\det W)^\dual$. A quick Chern class
computation yields that the cohomology is supported at $21$ points.
Upon closer inspection in Subsection~\ref{sec:requirements}, we will
see that two of these points are $p_1$ and $p_2$ again, leaving us
with $19$ other points which we denote as
\begin{equation}
  \label{eq:WdetW}
  \Hsupp\big( W^\dual \otimes \detinv W \big) = \big\{
  p_1, p_2, r_1, r_2, \dots, r_{19} 
  \big\}
  \,.
\end{equation}

Note that the points $q_i$, $s_j$ give the cohomology support of
bundles on $B_1$, and $p_k$, $r_l$ give the cohomology support of
bundles on $B_2$. For random values of the moduli, these two sets of
points will be disjoint, and all cohomology groups (including the
$\Ext^1$) vanish according to eq.~\eqref{eq:Hsupp}. Obviously, we want
to align some of the points to have extensions and a suitable matter
spectrum. Now the actual position of these points depends on complex
structure and vector bundle moduli, and one expects to be able to
align as many points as there are moduli to adjust. But actually
proving this would be cumbersome. Instead, we observe that one can
always adjust $3$ points by the way our Calabi-Yau threefold
$X=B_1\times_{\CP1}B_2$ is constructed. A priori, the \dP9 surfaces
$B_1\to\CP1$ and $B_2\to\CP1$ are elliptically fibered over two
different $\CP1$. In making the fiber product, one has to identify the
$\CP1$ bases. But one can always choose coordinates to fix $3$ points
on the sphere! Hence we can always pick a complex structure of $X$
such that
\begin{equation}
  q_1=p_1
  \,, \quad 
  q_2=p_2
  \,, \quad 
  s_1=r_1
  \,.
  \label{3.20}
\end{equation}
For this particular complex structure, 
\begin{equation}
  \Hsupp\big( L^{-4} \big)
  \cap
  \Hsupp\big( W^\dual \otimes \detinv W \big)
  = \big\{ p_1, p_2, r_1 \big\}
  \,,
\end{equation}
and therefore
\begin{equation}
  \Ext^1 \big( V_3, V_1 \big) = \C^3 \neq 0
  \,.
\end{equation}

Now we will show that with the identification eq.~\eqref{3.20} and
assuming that $p_1$ and $p_2$ appear in $\Hsupp(\wedge^2 W)$ with
multiplicity two (see Section~\ref{sec:explicit} for details), we can
make the number of the $SO(10)$ fundamentals 
\begin{equation}
  N_f=h^1\big( X, \wedge^2 V \big)
  = 8
  \,.
\end{equation}
This number satisfies the inequality eq.~\eqref{eq:Nfrange} for
$N_c=10$. In other words, we will prove that in the moduli space of
the complex structure and vector bundle there is at least one locus
where exactly $8$ fundamentals become light. The spectral cover
remains irreducible and the vector bundle remains smooth and stable
along this locus.  All other matter fields are massive and integrated
out. Moving slight away from this locus gives light masses to these
eight fundamental multiplets.  This is exactly what is need to satisfy
the criteria stated in Section~\ref{sec:breaking}. To move away from
this locus, for example, means to slightly separate $p_1$ from $q_1$
and $p_2$ from $q_2$. This is controlled by complex structure and/or
vector bundle moduli. In the next subsection, we will argue that the
$8$ fundamentals of interest receive a superpotential of the form
eq.~\eqref{2.1}.

To compute cohomology of $\wedge^2 V$ we have to relate it to
cohomology of $V_1$ and $V_3$. From the maps in eq.~\eqref{eq:Vext} we
can construct two exact sequences
\begin{equation}
  \begin{gathered}
    0 
    \longrightarrow
    \wedge^2 V_1
    \longrightarrow
    \wedge^2 V 
    \longrightarrow
    Q_1 
    \longrightarrow
    0
    \,, \\
    0 
    \longrightarrow
    Q_2 
    \longrightarrow
    \wedge^2 V 
    \longrightarrow
    \wedge^2 V_3 
    \longrightarrow
    0    
  \end{gathered}
\end{equation}
for some cokernel $Q_1$ and kernel $Q_2$. These two exact sequences
fit together into the commutative diagram 
\begin{equation}
  \label{eq:wedge2Vdiagram}
  \vcenter{\xymatrix{
      & 
      & 
      0 \ar[d]
      & 
      0 \ar[d]
      &
      \\
      0 \ar[r] &
      \wedge^2 \Vsheaf_1 \ar[r] \ar@{=}[d] &
      Q_2 \ar[r] \ar[d] &
      \Vsheaf_1 \otimes \Vsheaf_3 \ar[r] \ar[d] &
      0
      \\
      0 \ar[r] &
      \wedge^2 \Vsheaf_1 \ar[r] &
      \wedge^2 \Vsheaf \ar[r] \ar[d] &
      Q_1 \ar[r] \ar[d] &
      0
      \\
      & &
      \wedge^2 \Vsheaf_3 \ar@{=}[r] \ar[d] &
      \wedge^2 \Vsheaf_3 \ar[d] 
      \\
      & & 0 & 0
    }}
\end{equation}
with exact rows and columns. In our case $\wedge^2 V_1=0$ is the rank
$0$ vector bundle, since $V_1$ is a line bundle. Therefore the
commutative diagram simplifies to the short exact sequence
\begin{equation}
  0 
  \longrightarrow
  V_1 \otimes V_3 
  \longrightarrow
  \wedge^2 V 
  \longrightarrow
  \wedge^2 V_3 
  \longrightarrow
  0
  \,.
  \label{3.24}
\end{equation}
for $\wedge^2 V$. Using the definitions eqns.~\eqref{eq:V1def}
and~\eqref{eq:V1def}, the outer terms are
\begin{equation}
  \begin{split}
    \wedge^2 V_3 ~=&~
    \pi_1^\ast \big( L^2 \big) \otimes 
    \pi_2^\ast \big( \wedge^2 W \big)
    \,, \\
    V_1 \otimes V_3 ~=&~
    \pi_1^\ast\big( L^{-2} \big) \otimes 
    \pi_2^\ast \big(W \otimes \detinv W \big) 
    =
    \pi_1^\ast\big( L^{-2} \big) \otimes 
    \pi_2^\ast\big( \wedge^2 W^\dual \big)
    \,.
  \end{split}
\end{equation}
If we abbreviate
\begin{equation}
  \Fsheaf \eqdef
  \left[
    \pi_1^\ast \big( L^2 \big) \otimes 
    \pi_2^\ast\big( \wedge^2 W \big)
  \right]^\dual
  \,,
  \label{3.25}
\end{equation}
then the sequence eq.~\eqref{3.24} can be written as
\begin{equation}
  0 \to \Fsheaf \to \wedge^2 V \to \Fsheaf^\dual \to 0
  \,.
  \label{3.26}
\end{equation}
Now we can use the long exact sequence of cohomology to relate the
cohomology groups of $\wedge^2 V$ to the cohomology groups of
$\Fsheaf$. Since $\wedge^2 V$ is self-dual, Serre duality tells us
that $h^1(X, \wedge^2 V)=h^2(X, \wedge^2 V)$. Hence we can concentrate
on the part of the sequence involving $H^2(X, \wedge^2 V)$, which
reads
\begin{equation}
  \dots 
  \stackrel{\delta}{\longrightarrow}
  H^2\big(X, \Fsheaf\big)
  \longrightarrow 
  H^2\big(X, \wedge^2 V\big) 
  \longrightarrow
  H^2\big(X, \Fsheaf^\dual\big) 
  \longrightarrow
  H^3\big(X, \Fsheaf\big) = 0  
  \,,
  \label{3.27}
\end{equation}
where
\begin{equation}
  \delta : 
  H^1\big(X, \Fsheaf^\dual\big) \to H^2\big(X, \Fsheaf\big)
\label{3.28}
\end{equation}
is a coboundary map, which we can think of as a matrix with entries
depending on vector bundle moduli. It is determined by the chosen
extension class
\begin{equation}
  [\epsilon]
  \in
  \Ext^1\big(  V_3, V_1 \big) =
  H^1\big(X, V_1\otimes V_3^\dual \big) =
  H^1\Big(X, 
  \pi_1^\ast\big( L^{-4} \big) \otimes 
  \pi_2^\ast\big( W^\dual \otimes \detinv W \big)
  \Big)
  \,. 
  \label{3.29}
\end{equation}
The coboundary map eq.~\eqref{3.28} is simply multiplication by
$\epsilon$ followed by a suitable contraction of vector bundle
indices. The vector bundle extension, that is the cohomology of the
vector bundle $V_1\otimes V_3^\dual$, is supported at
\begin{equation}
  \Hsupp\big( L^{-4} \big)
  \cap
  \Hsupp\big( W^\dual \otimes \detinv W \big)
  = \big\{ p_1, p_2, r_1 \big\}
  \,,
\end{equation}
whereas the cohomology of $\Fsheaf^\dual$, $\Fsheaf$ is supported at
\begin{equation}
  \Hsupp\big( L^2 \big)
  \cap
  \Hsupp\big( \wedge^2 W \big)
  = \big\{ p_1, p_2 \big\}
  \,.
\end{equation}
Observe that the support of the extension class contains an additional
point over the support of the cohomology of $\Fsheaf^\dual$,
$\Fsheaf$. Hence, we can choose the extension class $[\epsilon]$ to be
localized at this additional point $r_1$, and we are going to do so in
the following. In that case the coboundary map $\delta$ is
automatically zero, and the sequence eq.~\eqref{3.27} becomes
\begin{equation}
  0 
  \longrightarrow
  H^2\big(X, \Fsheaf\big)
  \longrightarrow 
  H^2\big(X, \wedge^2 V\big) 
  \longrightarrow
  H^2\big(X, \Fsheaf^\dual\big) 
  \longrightarrow
  0  
  \,.
  \label{3.30}
\end{equation}
Therefore, 
\begin{equation}
  \label{eq:h2wedge2V}
  h^2\big(X,  \wedge^2 V\big) =
  h^2\big(X, \Fsheaf\big)+
  h^2\big(X, \Fsheaf^\dual\big)
  \,.
\end{equation}
The cohomology group $H^2(X, \Fsheaf^\dual)$ is straightforward to
calculate using the Leray spectral sequence, see also
eq.~\eqref{eq:Ucoh}. The answer is
\begin{equation}
  H^2\big(X, \Fsheaf^\dual) = 
  H^0\Big(\CP1,
  R^1\beta_{1\ast}\big( L^2 \big) \otimes 
  R^1\beta_{2\ast}\big( \wedge^2 W \big)
  \Big)
  \,.
  \label{3.32}
\end{equation}
We only have to be careful with the multiplicity of points in
$\Hsupp(\wedge^2 W)$. As discussed before, the push-down terms are
skyscraper sheaves supported at the points
\begin{equation}
  \label{3.33}
  \begin{split}
    R^1\beta_{1\ast} \big( L^2 \big) ~=&~
    \Osheaf_{q_1} \oplus 
    \Osheaf_{q_2} \oplus
    \Osheaf_{q_3}
    \,, 
    \\
    R^1\beta_{2\ast} \big( \wedge^2 W \big) ~=&~
    2 \Osheaf_{p_1} \oplus 
    2 \Osheaf_{p_2} \oplus
      \Osheaf_{p_3}
    \,,
  \end{split}  
\end{equation}
where $\Osheaf_{p}$ denotes the ``skyscraper'' sheaf which is a
one-dimensional vector space at $p$ at zero everywhere else.
Recalling our identifications eq.~\eqref{3.20}, we
obtain\footnote{Recall that $\Osheaf_p\otimes\Osheaf_p=\Osheaf_p$
  whereas $\Osheaf_p\otimes\Osheaf_q=0$ for $p\not=q$.}
\begin{equation}
  R^1\beta_{1\ast}\big( L^2 \big) \otimes 
  R^1\beta_{2\ast}\big( \wedge^2 W \big)
  =
  2{\cal O}_{p_1}\oplus 2{\cal O}_{p_2}.
  \label{3.34}
\end{equation}
Then using eq.~\eqref{3.32} it follows that
\begin{equation}
  h^2\big(X, \Fsheaf^\dual \big) = 
  h^0\big( \CP1, 2 \Osheaf_{p_1}\oplus 2 \Osheaf_{p_2}\big)  =
  4
  \,.
  \label{3.35}
\end{equation}
In exactly the same way one arrives at
\begin{equation}
  h^2\big(X, \Fsheaf\big) =
  4
  \,,
  \label{3.36}
\end{equation}
as well. Therefore, we find from eq.~\eqref{eq:h2wedge2V} that
\begin{equation}
  N_f = 
  h^2\big( X, \wedge^2 V \big) =
  h^1\big( X, \wedge^2 V \big) =
  8
  \,. 
  \label{3.37}
\end{equation}
Thus, our model indeed has $N_f=8$ massless fundamental multiplets,
satisfying the inequality eq.~\eqref{eq:Nfrange} for $N_c=10$. As
discussed earlier in this subsection we can give them small masses. In
the next subsection we will show that they have a superpotential of
the form eq.~\eqref{2.1}.

\subsection{Different Numbers of Flavors}
\label{sec:numbers}

In the following, we will always stick to the $N_f=8$ case in order to
make everything as explicit as possible. However, one can easily
construct similar bundles yielding different values for $N_f$. Let us
explore 
these possibilities.
\begin{itemize}
\item One simple change would be to deform one of the ordinary double
  points such that
  \begin{equation}
    \Hsupp\big(\wedge^2 W'\big) =\big\{2p_1, p_2, p_2', p_3\big\}
  \end{equation}
  for the new spectral curve $C'$. This is achieved by modifying the
  moduli of $W$.  In terms of the equations for the curve to be
  discussed in Section~\ref{sec:explicit}, this amounts to allowing
  the cubic $F_2$, eq.~\eqref{eq:pencil}, to be arbitrary.  Following
  exactly the same steps as in Subsection~\ref{sec:cohomology}, this
  yields $N_f=6$.
\item Similarly, by modifying the vector bundle moduli, it is easy to
  obtain any even $N_f$ less than six. For completeness, let us
  discuss this case even though it does not
  satisfy~\eqref{eq:Nfrange}. Consider the regime in the moduli space
  where
  \begin{equation}
    \Hsupp\big(\wedge^2 W'\big) =\big\{p_1, p_1', p_2, p_2', p_3\big\}
  \end{equation}
  A calculation analogous to the one performed in the previous
  subsection yields $N_f=4$. Now assuming that $\Hsupp(\wedge^2 W)$
  does not have any double points, let us move in the moduli space of
  complex structures so that the point $q_2 \in \Hsupp(L^2)$ gets
  separated from $p_2 \in \Hsupp(\wedge^2 W)$ and is not identified
  with any other point of $\Hsupp(\wedge^2 W)$. Then the analysis of
  the previous subsection yields $N_f=2$. Note that this separation
  also changes the possible non-trivial extensions. From
  eqns.~\eqref{3.18.5} and~\eqref{eq:WdetW} it follows that
  \begin{equation}
    \Ext^1(V_3, V_1) ={\mathbb C}^2.
    \label{n1}
  \end{equation}
  Separating further $q_1$ and $p_1$ makes the supports $\Hsupp(L^2)$
  and $\Hsupp(\wedge^2 W)$ completely disjoint.  This yields $N_f=0$.
  From eqns.~\eqref{3.18.5} and~\eqref{eq:WdetW} it follows that in this
  case
  \begin{equation}
    \Ext^1(V_3, V_1) ={\mathbb C}.
    \label{n2}
  \end{equation}
\item Another easy modification is to take a spectral curve $C''$ with
  one ordinary double points and one ordinary triple point. In terms
  of the intersection points $W''\big|_{f_1}$ this means (see
  eq.~\eqref{eq:Cfintersect}) that
  \begin{equation}
    S_1'' = \boxminus S_1'' = T_1''
    \,.
  \end{equation}
  To have enough parameters to adjust the spectral curve, one needs
  $C''\in\Gamma\OB2(\ell+2f)$, which we are free to choose. The only
  change to the matter spectrum of the resulting vector bundle is that
  now $N_f=10$.  Similarly, by considering the spectral cover of $W$
  to be in $\Gamma\OB2(\ell+nf)$ for greater values of $n$, it is
  possible to obtain greater values of $N_f$.
\item The fact that $N_f$ was always even so far is an artifact of the
  rank $3$ bundle $W$. This can be relaxed, for example, by
  constructing rank $5$ bundles as extension of a line bundle and a
  rank $4$ bundle $W'''$. The same trick of aligning points on the
  base $\CP1$ can then be used to adjust the matter spectrum. In this
  way, one can find odd $N_f$ lying in the range
  eq.~\eqref{eq:Nfrange}.
\end{itemize}
To conclude, we have shown that we can obtain the spectrum with
$N_f =0, \dots, 10, \dots$. Thus, in particular, 
we have shown that we can find any $N_f$ in the range~\eqref{eq:Nfrange}.


\subsection{The Superpotential}

Let us again consider the exact sequence eq.~\eqref{3.27}. Now let us
pick a generic extension $\epsilon$ instead of one supported only at
$r_1$. In that case the bundle extension is supported at the points
$p_1$, $p_2$ which support the cohomology of $\Fsheaf$. Then,
generically, the coboundary map $\delta$ becomes an isomorphism and
the exact sequence eq.~\eqref{3.27} becomes
\begin{equation}
  0 
  \longrightarrow 
  H^2\big(X, \wedge^2 V\big) 
  \stackrel{\sim}{\longrightarrow}
  H^2\big(X, \Fsheaf^\dual\big) 
  \longrightarrow
  0
  \,,
  \label{3.38}
\end{equation}
resulting in 
\begin{equation}
  N_f=
  h^1\big(X, \wedge^2 V\big)=
  h^2\big(X, \wedge^2 V\big)=
  h^2\big(X, \Fsheaf^\dual\big)=4
  \,.
  \label{3.39}
\end{equation}
In other words, turning on the vector bundle moduli parametrizing
$\Ext^1(V_3, V_1)$ we can remove half of $H^2(X, \wedge^2 V)$.
Similarly, turning on the anti-extension moduli coming from
$\Ext^1(V_1,V_3)$ we can remove the other half of $H^2(X, \wedge^2
V)$. This means that we have a superpotential that is quadratic in the
elements of $H^2(X, \wedge^2 V)$, giving mass to all fields at a
generic point in the moduli space.

Let us finish by giving a general explanation why coboundary maps
correspond to a cubic superpotential of the form in eq.~\eqref{2.1}.
Let the vector bundle $U$ be the extension of $U_1$ and $U_2$,
\begin{equation}
  0 
  \longrightarrow
  U_1 
  \longrightarrow
  U 
  \longrightarrow
  U_2 
  \longrightarrow
  0
  \,.
  \label{3.40}
\end{equation}
In the long exact sequence for the cohomology there is a coboundary
map
\begin{equation}
  \delta : H^1\big(X, U_2\big) 
  \to H^2\big(X, U_1\big)=H^1\big(X, U_1^\dual\big)^\dual. 
  \label{3.41}
\end{equation}
The map $\delta$ is a multiplication by a matrix $\epsilon$ of
differential forms parametrized by the vector bundle moduli. It is an
element of the extension group $[\epsilon]\in\Ext^1(U_2, U_1) = H^1(X,
U_1 \otimes U_2^\dual)$. Eq.~\eqref{3.41} says that the tensor product
$H^1(X, U_2) \otimes \Ext^1(U_2, U_1)$ defines an element in the dual
space $H^1(X, U_1^\dual)^\dual$. Elements of $H^1(X, U_1^\dual)^\dual$
can naturally be paired up with elements of $H^1(X, U_1^\dual)$ to
obtain a complex number. Thus, we can rewrite this map as
\begin{equation}
  H^1\big(X, U_2\big) \otimes 
  \Ext^1\big(U_2, U_1\big) \otimes 
  H^1\big(X, U_1^\dual\big) 
  \to \C
  \,.
  \label{3.42}
\end{equation}
Looking at the long exact sequence in cohomology, elements of $H^1(X,
U_2)$ label a quotient of $H^1(X, U)$. Similarly, $H^1(X, U_1^\dual)$
is a quotient of $H^1(X, U^\dual)$. In our case $\wedge^2 V$ is real
so both are some quotient of the same space $H^1(X, \wedge^2 V)$. The
corresponding four-dimensional fields $Q$ are in a real representation
of the low-energy gauge group. Finally, elements of $\Ext^1(U_2, U_1)$
are part of the vector bundle moduli~\cite{Buchbinder:2002ji,
  Braun:2005fk}.  Denote them as $\phi$.  Then the map
eq.~\eqref{3.42} is the algebraic version of the superpotential
eq.~\eqref{2.1}.


\section{The Geometry of the Spectral Cover}
\label{sec:explicit}

\subsection{Requirements}
\label{sec:requirements}

In this section, we will give a detailed explanations of why the
support of $\wedge^2 W$ can of the form eq.~\eqref{3.19}. Recall that,
as in eq.~\eqref{eq:FMC}, the restriction of $W$ to a generic fiber
$f$ is
\begin{equation}
  W\big|_f = 
  \Osheaf_f(C_1-o) \oplus 
  \Osheaf_f(C_2-o) \oplus 
  \Osheaf_f(C_3-o)
  \,,
\end{equation}
where the points $C_1$, $C_2$, $C_3$, and $o$ on $f$ are intersection
points with the spectral curve $C_W$ and the zero section $\sigma$,
\begin{equation}
  \big\{C_1,C_2,C_3\big\} = C_W \cdot f
  \,,\quad
  o = \sigma \cdot f
  \,.
\end{equation}
Tensor operations commute with the restriction, so we can simply write
down\footnote{The group law $\boxplus$ on the elliptic curve $f$
  satisfies $\Osheaf_f(p-o)\otimes\Osheaf_f(q-o) = \Osheaf_f\big(
  (p\boxplus q)-o\big)$.}
\begin{equation}
  \begin{split}
    \wedge^2 W\big|_f ~=&~ 
    \Osheaf_f(C_1\boxplus C_2-o) \oplus 
    \Osheaf_f(C_1\boxplus C_3-o) \oplus 
    \Osheaf_f(C_2\boxplus C_3-o)
    \,, \\
    \wedge^3 W\big|_f ~=&~ 
    \det W\big|_f =
    \Osheaf_f(C_1\boxplus C_2 \boxplus C_3 - o)
    \,.
  \end{split}
\end{equation}
Now we want a special spectral cover such that the cohomology support
$\Hsupp(\wedge^2 W)$ has a pair of points with multiplicity $2$. In
other words, on two special fibers 
\begin{equation}
  f_1 \eqdef \beta_2^{-1}(p_1)
  \,, \quad
  f_2 \eqdef \beta_2^{-1}(p_2)  
\end{equation}
we want (labeling the origin $o_i=\sigma \cdot f_i$)
\begin{equation}
  \begin{split}
    \wedge^2 W\big|_{f_1} ~=&~ 
    \Osheaf_{f_1} \oplus 
    \Osheaf_{f_1} \oplus 
    \Osheaf_{f_1}( a_1-o_1 )
    \,, \\
    \wedge^2 W\big|_{f_1} ~=&~ 
    \Osheaf_{f_1} \oplus 
    \Osheaf_{f_1} \oplus 
    \Osheaf_{f_1}( a_2-o_2 )
    \,.
  \end{split}  
\end{equation}
for some points $ a_i\in f_i-\{o_i\}$, $i=1,2$. In terms of the spectral
curve $C_W$, this means that we want
\begin{equation}
  \label{eq:Cfi}
  C_W \cdot f_i = \big\{ 2 S_i, T_i \big\}
\end{equation}
satisfying
\begin{equation}
  S_i \boxplus T_i = o_i
  \,, \quad
  S_i \boxplus S_i = a_i \not= o_i
  \,.
\end{equation}
Note that there are two ways to achieve intersections with
multiplicities as in eq.~\eqref{eq:Cfi}. Deforming the fiber away from
$f_i$, the intersection points with $C_W$ have to split up into $3$
distinct points. This triple can have a monodromy as one moves around
$f_i$, or it can have no monodromy. In the first case the spectral
curve $C_W$ has a branch point, while in the second case the spectral
curve has an ordinary double point. The corresponding spectral curve
for $\wedge^2 W$ has a worse singularity in the first case, and again
an ordinary double point in the second case. Now an ordinary double
point is simply a transverse intersection of two different sheets of
$C_W$. While technically called a singularity, it does not change
anything for the spectral cover construction. Hence, we will demand
that the points $S_1$ and $S_2$ are ordinary double points of $C_W$.
Such a spectral curve would yield
\begin{subequations}
  \begin{align}
    W\big|_{f_i} ~=&~ 
    \Osheaf_{f_i}(S_i-o_i) \oplus 
    \Osheaf_{f_i}(S_i-o_i) \oplus 
    \Osheaf_{f_i}(T_i-o_i)
    \,, \\
    W^\dual\big|_{f_i} ~=&~ 
    \Osheaf_{f_i}(\boxminus S_i-o_i) \oplus 
    \Osheaf_{f_i}(\boxminus S_i-o_i) \oplus 
    \Osheaf_{f_i}(\boxminus T_i-o_i)
    \,, \\
    \wedge^2 W\big|_{f_i} ~=&~ 
    \Osheaf_{f_i} \oplus 
    \Osheaf_{f_i} \oplus 
    \Osheaf_{f_i}( S_i\boxplus S_i-o_i )
    \,, \\   
    \det W\big|_{f_i} ~=&~ 
    \Osheaf_{f_i}( S_i\boxplus S_i\boxplus T_i-o_i )
    = 
    \Osheaf_{f_i}( S_i -o_i )
    \,, \\ 
    \label{eq:WdetWfi}
    \left.\Big( W^\dual \otimes \detinv W\Big)\right|_{f_i} ~=&~ 
    \Osheaf_{f_i}(\boxminus S_i\boxminus S_i-o_i) \oplus 
    \Osheaf_{f_i}(\boxminus S_i\boxminus S_i-o_i) \oplus 
    \Osheaf_{f_i}    
    \,,    
  \end{align}
\end{subequations}
as desired. Note that the last equation, eq.~\eqref{eq:WdetWfi}, tells
us that the cohomology of $W^\dual \otimes \detinv W$ is also
supported at $p_1$ and $p_2$, which we announced previously in
eq.~\eqref{eq:WdetW}.

To summarize, we require that our spectral curve satisfies
\begin{itemize}
\item $C_W \in \Gamma \OB2\big( \ell +f \big) $, see
  eq.~\eqref{eq:CWdef}.
\item $C_W$ has $2$ ordinary double points $S_1$ and $S_2$ in two
  different fibers $f_1$ and $f_2$, which we take to be non-degenerate
  elliptic curves for simplicity.
\item Then there are two more points $T_1$, $T_2$ satisfying
  \begin{equation}
    \label{eq:Cfintersect}
    f_i \cdot C_W = \{ 2 S_i, T_i \}
    \,.
  \end{equation}
  With respect to the group law on the elliptic curves, we require
  that
  \begin{equation}
    S_i \boxplus T_i = \sigma \cdot f_i
  \end{equation}
\item The double points do not coincide with the origin, that is
  \begin{equation}
    S_i \not= \sigma \cdot f_i 
    \quad\Leftrightarrow\quad 
    T_i \not= \sigma \cdot f_i 
    \quad\Leftrightarrow\quad 
    \det W\big|_{f_i}\not=\Osheaf_{f_i} 
    \,.
  \end{equation}
\end{itemize}

\subsection{The Pencil of Cubics}
\label{sec:pencil}

So far, we assumed the existence of a suitable spectral curve $C_W$ in
order to construct our vector bundle. Given that the surface $B_2$ has
$10$ and the curve $C_W$ has $5$ parameters, it is very plausible that
some choice of \dP9 surface and curve actually satisfies the
requirements laid out in Subsection~\ref{sec:requirements}. The
purpose of this section is to write an explicit spectral curve and
show that it satisfies all requirements.  This will establish the
existence of curve $C_W$ and, hence, of our vector bundle.

First, we have to specify the actual \dP9 surface $B_2$. We define it
as a ``Pencil of Cubics'', that is, a bi-degree $(3,1)$ hypersurface
in $\CP2\times\CP1$. In the following, we are going to use coordinates
$[x:y:z]$ for the coordinates on $\CP2$ and $[u:v]$ for $\CP1$. Define
the two cubics
\begin{equation}
  \begin{split}
    F_1(x,y,z) ~=&~  
    \left( x-y \right)  \left( x-z \right)  \left( x+z \right) 
    +z \left( x^2+y^2-z^2-2\,yx-4\,zx+5\,yz \right) 
    \\
    F_2(x,y,z) ~=&~ 
    \left( x-z \right)  \left( x-y \right)  \left( x+y \right) 
    +y \left( x^2+z^2-y^2-2\,zx-4\,yx+5\,yz \right) \\
    =&~ F_1(x,z,y)
    \,,
  \end{split}
\end{equation}
then 
\begin{equation}
  \label{eq:pencil}
  P\big( x,y,z;\,u,v \big) = 
  u\, F_1(x,y,z) + v\, F_2(x,y,z)
\end{equation}
is the desired equation. We define
\begin{equation}
  B_2 = \Big\{ P=0 \Big\} \subset \CP2\times\CP1
  \,.
\end{equation}
The elliptic fibration $\beta_2:B_2\to\CP1$ is just the projection on
the second factor, and we write
\begin{equation}
  f_{[u_0:v_0]} 
  \eqdef \beta_2^{-1}\big( [u_0:v_0] \big) 
\end{equation}
for the fiber over $[u_0:v_0]\in \CP1$. The discriminant locus of the
elliptic fibration is
\begin{multline}
  \Delta(P) = 
  \frac {25}{16}\, 
  \Big( 
    131\,{v}^{10}
    +5774\,u{v}^{9}
    -94185\,{u}^{2}{v}^{8}
    \\
    -2553672\,{u}^{3}{v}^{7}
    +26073510\,{u}^{4}{v}^{6}
    -49632012\,{u}^{5}{v}^{5}
    +26073510\,{u}^{6}{v}^{4}
    \\
    -2553672\,{u}^{7}{v}^{3}
    -94185\,{u}^{8}{v}^{2}
    +5774\,{u}^{9}v
    +131\,{u}^{10} 
    \Big)  
  \big( u+v \big)^2
\end{multline}
We observe that $B_2$ is a smooth surface, but of course some fibers
of the elliptic fibration are degenerate. More precisely, $B_2$ has
$10 I_1$ and $1 I_2$ singular Kodaira fibers, none of which lie over
the two points $[u:v]=[1:0],[0:1]$.

Note that the point $[2:1:1]\in \CP2$ is a basepoint of the pencil of
cubics (of mul\-ti\-plic\-i\-ty $1$). That is, 
\begin{equation}
  F_1(2,1,1)=0=F_2(2,1,1)
  \,,\quad
  -10=
  \left.\frac{\partial F_1}{\partial y}\right|_{(2,1,1)}
  \not=
  \left.\frac{\partial F_2}{\partial y}\right|_{(2,1,1)}
  = 0
  \,.
\end{equation}
Such a basepoint defines a section of the elliptic fibration, which we
declare to be the zero section
\begin{equation}
  e_9 ~\eqdef~
  \Big\{ \big( [2:1:1],[u:v] \big) 
  \,\Big|\, [u:v]\in \CP1 \Big\} \subset B_2
\end{equation}

\subsection{The Spectral Curve}
\label{sec:curve}

Having fixed the \dP9 surface $B_2$, we are now going to pick a curve
$C_W$ on it. For that, we define the equation
\begin{equation}
  Q\big( x,y,z;\,u,v \big) = 
  vz-uy
\end{equation}
on $B_2\subset \CP2\times\CP1$. Its zero locus will be the curve
\begin{equation}
  C_W ~\eqdef~ \Big\{ Q=0 \Big\} \subset B_2
  \,.
\end{equation}
Clearly, $C$ is a $3$-section of the elliptic fibration since its
intersection with the fiber over $[u_0:v_0]\in \CP1$ is given by the
cubic equation 
\begin{equation}
  C_W|_{f_{[u_0:v_0]}} = 
  \Big\{ 
  P(x,y,z;\,u_0,v_0)=0=Q(x,y,z;\,u_0,v_0)
  \Big\}
  \,.
\end{equation}
Note that a degree $(1,0)$ equation is, by definition, the hyperplane
section of $\CP2$, which is the homology class
\begin{equation}
  \Big[ \{x=0\} \Big] 
  = 
  \Big[ \{y=0\} \Big] 
  = 
  \Big[ \{z=0\} \Big] 
  = 
  \ell \in H_2\big( B_2,\Z \big)
  \,.
\end{equation}
Similarly, a degree $(0,1)$ equation cuts out one elliptic fiber of
$B_2$,
\begin{equation}
  \Big[ \{u=0\} \Big] 
  = 
  \Big[ \{v=0\} \Big] 
  = 
  f \in H_2\big( B_2,\Z \big)
  \,.
\end{equation}
Therefore
\begin{equation}
  C_W \in \Gamma \OB2\big( \ell+f \big)
  \quad \Rightarrow \quad
  [C_W] = \ell + f \in H_2\big( B_2 \big)
  \,.
\end{equation}
Computing the monodromies around branch points of $C_W$, we find that
it is an irreducible curve.

The curve $C_W$ is singular since having two ``ordinary double points''
was part of the requirements. These two points are
\begin{equation}
  \begin{split}
    S_1 ~\eqdef&~
    \Big( [0:0:1], [1:0] \Big)
    \quad \in f_{[1:0]}
    \,,
    \\
    S_2 ~\eqdef&~
    \Big( [0:1:0], [0:1] \Big)
    \quad \in f_{[0:1]}
    \,.
  \end{split}
\end{equation}
Since each fiber contains $3$ points of $C_W$ (counted with
multiplicity), there is another point in $f_{[1:0]}$ and $f_{[1:0]}$,
respectively. They are smooth points of $C_W$, and we label them
\begin{equation}
  \begin{split}
    T_1 ~\eqdef&~
    \Big( [1:0:1], [1:0] \Big)
    \quad \in f_{[1:0]}
    \,,
    \\
    T_2 ~\eqdef&~
    \Big( [1:1:0], [0:1] \Big)
    \quad \in f_{[0:1]}
    \,.
  \end{split}
\end{equation}
Apart from $S_1$ and $S_2$, there are no other singularities. As a
$3$-sheeted cover of the base $\CP1$ there are $6$ branch points in
other fibers,
\begin{figure}[htbp]
  \centering
  \input{threesheetedcover.pstex_t}
  \caption{The $3$-section $C_W$.}
  \label{fig:curve}
\end{figure}
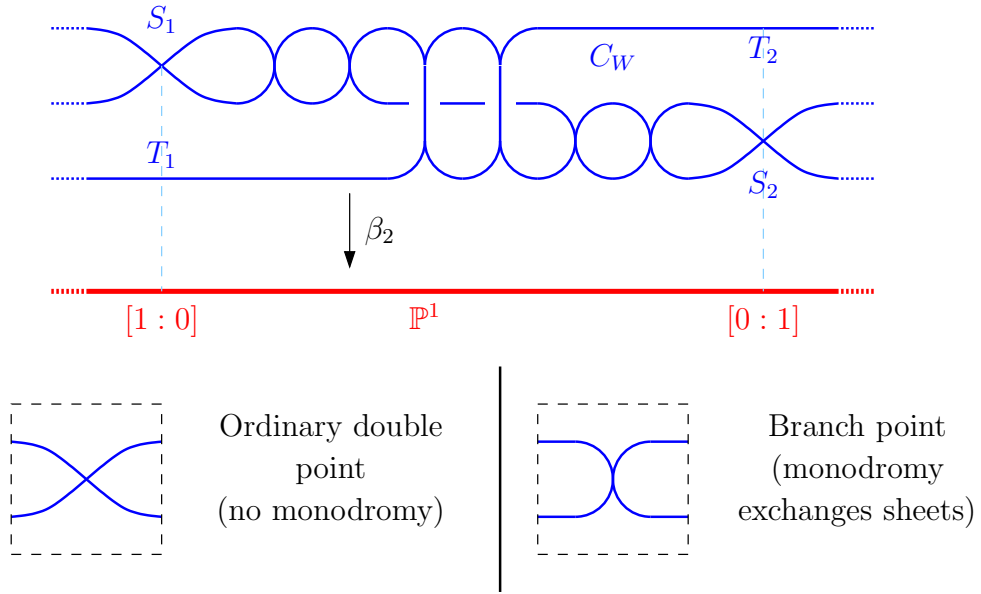
this is depicted in Figure~\ref{fig:curve}. 

It remains to show that 
\begin{equation}
  S_1 \boxplus T_1 = 0
  \,, \quad
  S_2 \boxplus T_2 = 0  
\end{equation}
in the group law on the respective fibers, then the curve $C_W$
satisfies all requirements. Note that everything so far was by
construction symmetric under the exchange
\begin{equation}
  \Big( [x:y:z],\,[u:v] \Big)  
  ~\leftrightarrow~
  \Big( [x:z:y],\,[v:u] \Big)    
\end{equation}
Because of this symmetry, it suffices to show that $S_2\boxplus
T_2=0$. The elliptic curve $f_{[0:1]}\subset \CP2$ is given by the
cubic
\begin{equation}
  \label{eq:P01cubic}
  P\big(x,y,z;\,0,1) =
  {x}^{3}-{z}^{3}
  -5\,x{z}^{2}-{x}^{2}y+6\,y{z}^{2}+{x}^{2}z+{y}^{2}z
  -2\,xyz
\end{equation}
with origin
\begin{equation}
  e_9 \cap f_{[0:1]} = [2:1:1] \in \CP2
\end{equation}
To bring the cubic into Weierstrass form we have to do a birational
transformation of the $\CP2$. Specifically, we choose new projective
coordinates $[X:Y:Z]$ via
\begin{equation}
  \begin{split}
    x ~=&~
    2\, \left( X-Z \right)  \left( 2\,X-7\,Z \right) 
    \,,
    \\
    y ~=&~
    2\,{X}^{2}-34\,X Z^2+57\,Z^3+5\,Y Z^2
    \,,
    \\
    z ~=&~
    2\, \left( X-Z \right)^{2}
    \,,
  \end{split}
\end{equation}
which maps the chosen origin to\footnote{$[0:1:0]$ is the origin in
  the Weierstrass form of a cubic.} $[0:1:0]$ in the new coordinates.
Substituting into eq.~\eqref{eq:P01cubic} we find
\begin{equation}
  P\big(X,Y,Z;\,0,1) =
  -50\,Z \left( X-Z \right) ^{2} 
  \left( -{Y}^{2}Z + 4\,{X}^{3}-52\,X{Z}^{2}+73\,{Z}^{3} \right)  
\end{equation}
Hence, the Weierstrass form of our elliptic curve is
\begin{equation}
  {Y}^{2}Z = 4\,{X}^{3}-52\,X{Z}^{2}+73\,{Z}^{3}
\end{equation}
The coordinates of the points $S_2$ and $T_2$ turn out to be at the
locus where the birational transformation is not defined, but one can
still find their values by continuity.
\begin{table}
  \centering
  \renewcommand{\arraystretch}{1.3}
  \begin{tabular}{c|ccc}
    $\displaystyle \CP2 $& 
    $\displaystyle S_1 $& 
    $\displaystyle T_2 $& 
    $\displaystyle e_9 \cap f_{[0:1]}  $\\
    \hline
    $\displaystyle [x:y:z]$ coordinates &
    $\displaystyle [0:1:0]   $&
    $\displaystyle [1:1:0]   $&
    $\displaystyle [2:1:1]   $\\
    $\displaystyle [X:Y:Z]$ coordinates &
    $\displaystyle [1: 5:1]   $&
    $\displaystyle [1:-5:1]   $&
    $\displaystyle [0:1:0]   $
  \end{tabular}
  \caption{Coordinate transformation to Weierstrass form}
  \label{tab:coordinates}
\end{table}
The new coordinates of the relevant points are listed in
Table~\ref{tab:coordinates}. Recall that the inverse in the group law
of the cubic has a particularly nice form for a cubic in Weierstrass
form, it is
\begin{equation}
  \boxminus [X:Y:Z] = [X:-Y:Z]
  \,.
\end{equation}
Hence, we immediately realize that
\begin{equation}
  \boxminus S_2 = T_2
  \quad \Leftrightarrow \quad
  S_2 \boxplus T_2 = 0
  \,,
\end{equation}
as required.


\section{Conclusion and Further Directions}
\label{sec:conclusion}


In this paper, we addressed the question of realizing dynamical SUSY
breaking~\cite{Intriligator:2006dd} in heterotic model building.  We
discussed how quadratic superpotentials for matter fields arise in
heterotic compactifications. The mass of these fields depends on the
complex structure and vector bundle moduli. Thus, by moving in the
moduli space, we can make some of the matter fields either very light
or very heavy. From an algebraic geometry viewpoint, this means that
the dimension of various cohomology groups associated to the number of
matter particles jumps as we move in the moduli space.  We present a
stable, holomorphic hidden sector bundle satisfying the criteria for
dynamical SUSY breaking. 
The main example studied in this paper is 
$SO(10)$ SQCD with
$N_f=8$ fundamental fields. All other matter fields are heavy and
integrated out. We give a detailed analysis showing that there is a
locus in the moduli space where exactly eight fundamentals become
massless whereas all other matter is massive.  Moving slightly away
from this locus is equivalent to generate the superpotential
eq.~\eqref{2.1}. Hidden sectors for different values of $N_f$ can be
constructed analogously. This is discussed in subsection 5.6. 
In particular, it is shown that it is possible to 
obtain $SO(10)$ SQCD with any number of fundamental fields in the 
range~\eqref{eq:Nfrange}

Let us briefly discuss various generalizations of these results. One
natural direction is to construct the hidden sector breaking
supersymmetry in realistic standard model
compactifications~\cite{Braun:2005ux, Braun:2005bw, Braun:2005zv,
  Braun:2005nv, Braun:2006ae, Bouchard:2005ag}. That is, in addition
to the sector whose particle spectrum is that of a supersymmetric
standard model, to put a hidden sector vector bundle (presumably one
without Wilson lines) that will lead to one of the theories studied
in~\cite{Intriligator:2006dd}. Another direction would be to
understand the F-theory dual~\cite{Vafa:1996xn, Friedman:1997yq} of a
model studied in this paper.  The F-theory dual space is a Calabi-Yau
fourfold $Y$ and the matter is supposed to be localized on
intersecting $D7$-branes wrapping four-cycles of $Y$.  The moduli of
the heterotic vector bundle will be mapped to certain geometric moduli
of $Y$. Thus, giving mass to the fundamentals by means of the
superpotential eq.~\eqref{2.1} on the F-theory side will have a
geometric interpretation as brane separation. Having this
interpretation, it might be easier to understand the location of the
loci where the right number of the fundamental multiplets receive a
small mass. Another possible advantage of it is that it could be
easier to understand under what conditions the moduli controlling the
masses of the fundamentals can be stabilized in a regime of interest.
Unfortunately, it is not yet known how the heterotic/F-theory duality
map acts on the spectrum.


\section{Acknowledgments}

The authors would like to thank Ken Intriligator, Tony Pantev, and
Nathan Seiberg for valuable discussions and explanations. This
research was supported in part by the Department of Physics and the
Math/Physics Research Group at the University of Pennsylvania under
cooperative research agreement DE-FG02-95ER40893 with the
U.~S.~Department of Energy and an NSF Focused Research Grant
DMS0139799 for ``The Geometry of Superstrings''. The work of E.I.B. is
supported by NSF grant PHY-0503584.


\appendix

\section{\texorpdfstring{The Support of $\mathbf{L^k}$}{The Support of L}}
\label{sec:Lsupp}

Let $\Lsheaf=\Osheaf_B(s_1-s_2)$ be a line bundle on a \dP9 surface
$B$, and $s_1$ and $s_2$ be two sections that do not intersect. For
example, one can use $s_1=e_1$ and $s_2=e_9$ where $e_1$ and $e_9$ are
two out of the nine exceptional divisors from the blow-up. Thus both
$e_1$ and $e_9$ are isomorphic to $\CP1$.  We will denote by $\beta$
the projection of $B$ to the base $\CP1$. Via the Leray spectral
sequence, the cohomology of $\Lsheaf$ is determined by the cohomology
(on $\CP1$) with coefficients in either $\beta_\ast \Lsheaf$ or
$R^1\beta_\ast \Lsheaf$. In particular, as the base is
$1$-dimensional one obtains
\begin{equation}
\begin{split}
  H^0\big( B,\Lsheaf \big) ~=&~
  H^0\big( \CP1,\beta_\ast \Lsheaf \big)
  \,, \\
  H^1\big( B,\Lsheaf \big) ~=&~
  H^0\big( \CP1,R^1\beta_\ast \Lsheaf \big) \oplus
  H^1\big( \CP1,\beta_\ast \Lsheaf \big)
  \,, \\
  H^2\big( B,\Lsheaf \big) ~=&~
  H^1\big( \CP1,R^1\beta_\ast \Lsheaf \big)
  \,.
\end{split}
\end{equation}
Since $s_1$ and $s_2$ do not intersect, the restriction of $\Lsheaf$
to any fiber gives a non-trivial line bundle of degree zero on
elliptic curve, see Section~\ref{sec:jumping}. Such a line bundle (on
a fiber) has no cohomology, that is, all cohomology groups vanish.
Therefore, both $\beta_\ast \Lsheaf$ or $R^1\beta_\ast \Lsheaf$ are
the zero sheaf
\begin{equation}
  \beta_\ast \Lsheaf =R^1\beta_\ast \Lsheaf
  =0
  \,.
  \label{A1}
\end{equation}
This means that in turn all cohomology groups (on $B$) of $\Lsheaf$
vanish,
\begin{equation}
  H^\ast\big( B,\Lsheaf \big) =0
  \,.
\end{equation}
Let us now consider $\Lsheaf^2 =\Osheaf_B(2s_1-2s_2)$.  As we will see
below, for this line bundle $\beta_\ast \Lsheaf^2$ is still zero.
However, $R^1\beta_\ast \Lsheaf^2$ is not zero. Instead, it is a
torsion sheaf. According to the Leray spectral sequence,
\begin{equation}
  H^1\big(B, \Lsheaf^2\big) =
  H^0\big(\CP1, R^1\beta_\ast \Lsheaf^2\big)
  \,. 
  \label{A2}
\end{equation}
The right hand side is just the number of points at which
$R^1\beta_\ast \Lsheaf^2$ is supported. Let us calculate this number.
To do so we will be using the following standard exact
sequence~\cite{MR507725}. Let $D$ be any effective divisor on a
manifold $Z$, then the sequence
\begin{equation}
  0
  \longrightarrow
  \Osheaf_Z(-D) 
  \longrightarrow
  \Osheaf_Z 
  \longrightarrow
  \Osheaf_D
  \longrightarrow
  0
  \,. 
  \label{A3}
\end{equation}
is exact. The first map $\Osheaf_Z(-D) \to \Osheaf_Z$ is 
multiplication by a global section of $\Osheaf_Z(D)$ which vanishes
exactly at $D$. The second map $\Osheaf_Z\to\Osheaf_D$ is simply
the restriction to $D\subset Z$.

Let us apply it to the case when $Z \simeq \CP1$ and $D \simeq
\ptset$. Then we have
\begin{equation}
  0 
  \longrightarrow
  \OP(-1) 
  \longrightarrow
  \OP 
  \longrightarrow
  \Osheaf_p 
  \longrightarrow
  0
  \,. 
  \label{A4}
\end{equation}
The cokernel $\Osheaf_p$ is the skyscraper sheaf supported at a point
$p$. This sequence can easily be generalized for the case of $n$
points, and the cokernel of the inclusion map 
\begin{equation}
  i: \OP(-n) \to \OP 
\end{equation}
is a skyscraper sheaf supported at $n$ points. The detailed location
of these points depends on the map $i$ which is multiplication by a
global section of $\Osheaf(n)$. Any such section has $n$ zeroes which
are the support of the cokernel of $i$. Conversely, for any $n$ points
there is such a (unique up to an overall constant) section vanishing
at the $n$ points.

Now let us apply the short exact sequence eq.~\eqref{A2} to the bundle
$\Lsheaf^2$. First we consider the sequence
\begin{equation}
  0
  \longrightarrow
  \Osheaf_B (s_1-s_2)
  \longrightarrow
  \Osheaf_B(2s_1-s_2)
  \longrightarrow
  \Osheaf_{s_1}( 2s_1 \cdot s_1)
  \longrightarrow
  0
  \,.
  \label{A5}
\end{equation}
The bundle $\Osheaf_{s_1}( 2s_1 \cdot s_1)$ is a bundle on $s_1 \sim
\CP1$ of degree $-2 = 2 s_1^2$. This short exact sequence on $B$ leads
to a long exact sequence on $\CP1$ of the direct images
\begin{equation}
  0 
  \longrightarrow
  \beta_\ast\Osheaf_B (s_1-s_2)
  \longrightarrow
  \beta_\ast\Osheaf_B(2s_1-s_2) \to O(-2) 
  \longrightarrow
  R^1 \beta_\ast \Osheaf_B (s_1-s_2) 
  \longrightarrow
  \cdots
  \,. 
\label{A6}
\end{equation}
We have shown above that $\beta_\ast\Osheaf_B (s_1-s_2)=R^1
\beta_\ast\Osheaf_B (s_1-s_2)=0$. Therefore,
\begin{equation}
  \label{A7}
  \begin{split}
    \beta_\ast\Osheaf_B(2s_1-s_2) ~=&~
    \OP(-2)
    \,,  \\
    R^1\beta_\ast\Osheaf_B(2s_1-s_2) ~=&~ 0
  \end{split}
\end{equation}
Finally, consider the sequence
\begin{equation}
  0
  \longrightarrow
  \Osheaf_B (2s_1-2s_2) 
  \longrightarrow
  \Osheaf_B(2s_1-s_2) 
  \longrightarrow
  \OP(1) 
  \longrightarrow
  0
  \,, 
  \label{A8}
\end{equation}
where we used the intersection numbers eq.~\eqref{3.1.2} already.
Inserting eq.~\eqref{A7}, we find direct images
\begin{equation}
  0 
  \longrightarrow
  \OP(-2)
  \longrightarrow
  \OP(1) 
  \longrightarrow
  R^1 \beta_\ast \big( \Lsheaf^2 \big)
  \longrightarrow
  0
  \,. 
  \label{A9}
\end{equation}
From the above discussion it follows that $R^1 \beta_\ast
\big(\Lsheaf^2\big)$ is the skyscraper sheaf supported at $3$ points.
In Section~\ref{sec:model}, we denoted these points by $q_1, q_2,
q_3$.  Similarly, one can show that the sheaf $R^1 \beta_\ast
\big(\Lsheaf^3\big)$ is supported at $8$ points and the sheaf $R^1
\beta_\ast\big( \Lsheaf^4 \big)$ is supported at $15$ points. It is not
hard to show that these $15$ points contain the points $q_1$, $q_2$,
and $q_3$ each with multiplicity one.

\section{Derived Tensor Products}
\label{sec:derived}

Consider the case of a bundle on $X=B_1\times_{\CP1}B_2$ constructed
as
\begin{equation}
  U = 
  \pi_1^\ast \big( U_1 \big) \otimes 
  \pi_2^\ast \big( U_2 \big) 
  \,,
\end{equation}
as we are using throughout this paper. Moreover, let the bundles $U_i$
on $B_i$ be semistable of fiber-degree zero. For ease of presentation,
we assume that their direct image contains only a single skyscraper
sheaf, that is,
\begin{align}
  \beta_{1\ast}\big(U_1\big)   ~=&~ 
  0 
  \,,
  & 
  R^1 \beta_{1\ast}\big(U_1\big)   ~=&~ 
  \Osheaf_p
  \,,
  \\
  \beta_{2\ast}\big(U_2\big)   ~=&~ 
  0 
  \,,
  &
  R^1 \beta_{2\ast}\big(U_2\big)   ~=&~ 
  \Osheaf_q
\end{align}
for two points $p,q\in \CP1$. To compute the cohomology we can apply
the Leray spectral sequence, either pushing down to $B_1$ or to
$B_2$,
\begin{equation}
  \begin{split}
    H^i\big( X,U \big) ~=&~
    \bigoplus_{n+m=i}
    H^n\Big( 
    U_1 \otimes
    \beta_1^\ast \circ R^m \beta_{2\ast} \big(U_2\big) 
    \Big)
    = \\ =&~
    \bigoplus_{n+m=i}
    H^n\Big( 
    \beta_2^\ast \circ R^m \beta_{1\ast} \big(U_1\big) 
    \otimes U_2 \Big)
  \end{split}
\end{equation}
However, a problem arises when one attempts to push either term
further down to $\CP1$. Because the $R^1\beta_{i\ast}(U_i)$ is not a
vector bundle we cannot simply apply the projection formula, and
\begin{equation}
  \begin{split}
    R^n \beta_{1\ast} \big(U_1\big) 
    \otimes 
    R^m \beta_{2\ast} \big(U_2\big) 
    ~\not=&~
    R^n\beta_{1\ast}\Big(
    U_1 \otimes
    \Big[
    \beta_1^\ast \circ R^m \beta_{2\ast} \big(U_2\big) 
    \Big]
    \Big)
    \not= \\ \not=&~
    R^m\beta_{2\ast}\Big(
    \Big[
    \beta_2^\ast \circ R^n \beta_{1\ast} \big(U_1\big) 
    \Big]
    \otimes U_2
    \Big)
  \end{split}
\end{equation}
in general. The solution to this problem is well-known, one has to
work in the derived category. That is, the tensor product has to be
replaced by the derived tensor product, and we have to take the
hypercohomology of the resulting complexes. Fortunately, this is
relatively easy for skyscraper sheaves on $\CP1$. Their derived tensor
product is simply
\begin{equation}
  \Osheaf_p 
  ~\underset{L}{\otimes}~
  \Osheaf_q
  =
  \begin{cases}
    0
    \,, & 
    p\not=q
    \,, \\
    \Osheaf_p \oplus \Osheaf_p[-1] 
    \,, & 
    p=q
    \,.
  \end{cases}
\end{equation}
Therefore, if $p\not=q$ then
\begin{equation}
  H^i\big( X,U \big) =0 
  \,,
\end{equation}
whereas if $p=q$ then
\begin{equation}
  \lrstack[b]{
    H^i\big( X,U \big) =
    H^i\Big(
    R^\ast \beta_{1\ast} \big(U_1\big) 
    ~\underset{L}{\otimes}~
    R^\ast \beta_{2\ast} \big(U_2\big)   
    \Big) =
  }{
    =
    H^i\Big( \Osheaf_p[1] ~\underset{L}{\otimes}~ \Osheaf_q[1]  \Big) =
    H^i\Big( \Osheaf_p[2] \oplus \Osheaf_p[1]  \Big) 
  }
  \simeq
  \begin{cases}
    0  & i=3 \,, \\
    \C & i=2 \,, \\
    \C & i=1 \,, \\
    0  & i=0 \,.
  \end{cases}  
\end{equation}
Notice that we could have used the ordinary tensor product and
cohomology as long as we are only computing $H^2(X,U)$. This is
precisely what we did in Section~\ref{sec:model}, and it is justified
through the above computation.


\bibliographystyle{JHEP} \renewcommand{\refname}{Bibliography}
\addcontentsline{toc}{section}{Bibliography} \bibliography{susy}

\end{document}

%% file: threesheetedcover.pstex_t
\begin{picture}(0,0)%
\includegraphics{threesheetedcover.pstex}%
\end{picture}%
\setlength{\unitlength}{4144sp}%
\begingroup\makeatletter\ifx\SetFigFont\undefined%
\gdef\SetFigFont#1#2#3#4#5{%
  \reset@font\fontsize{#1}{#2pt}%
  \fontfamily{#3}\fontseries{#4}\fontshape{#5}%
  \selectfont}%
\fi\endgroup%
\begin{picture}(5230,3553)(204,-3008)
\put(1126,389){\makebox(0,0)[b]{\smash{{\SetFigFont{12}{14.4}{\rmdefault}{\mddefault}{\updefault}{\color[rgb]{0,0,1}$S_1$}%
}}}}
\put(2701,-1411){\makebox(0,0)[b]{\smash{{\SetFigFont{12}{14.4}{\rmdefault}{\mddefault}{\updefault}{\color[rgb]{1,0,0}$\CP1$}%
}}}}
\put(3826,164){\makebox(0,0)[b]{\smash{{\SetFigFont{12}{14.4}{\rmdefault}{\mddefault}{\updefault}{\color[rgb]{0,0,1}$C_W$}%
}}}}
\put(1126,-421){\makebox(0,0)[b]{\smash{{\SetFigFont{12}{14.4}{\rmdefault}{\mddefault}{\updefault}{\color[rgb]{0,0,1}$T_1$}%
}}}}
\put(4726,209){\makebox(0,0)[b]{\smash{{\SetFigFont{12}{14.4}{\rmdefault}{\mddefault}{\updefault}{\color[rgb]{0,0,1}$T_2$}%
}}}}
\put(4726,-601){\makebox(0,0)[b]{\smash{{\SetFigFont{12}{14.4}{\rmdefault}{\mddefault}{\updefault}{\color[rgb]{0,0,1}$S_2$}%
}}}}
\put(1126,-1411){\makebox(0,0)[b]{\smash{{\SetFigFont{12}{14.4}{\rmdefault}{\mddefault}{\updefault}{\color[rgb]{1,0,0}$[1:0]$}%
}}}}
\put(4726,-1411){\makebox(0,0)[b]{\smash{{\SetFigFont{12}{14.4}{\rmdefault}{\mddefault}{\updefault}{\color[rgb]{1,0,0}$[0:1]$}%
}}}}
\put(2341,-871){\makebox(0,0)[lb]{\smash{{\SetFigFont{12}{14.4}{\rmdefault}{\mddefault}{\updefault}{\color[rgb]{0,0,0}$\beta_2$}%
}}}}
\put(1351,-2311){\makebox(0,0)[lb]{\smash{{\SetFigFont{12}{14.4}{\rmdefault}{\mddefault}{\updefault}{\color[rgb]{0,0,0}\parbox{3.5cm}{\begin{center}Ordinary double point\\(no monodromy)\end{center}}}%
}}}}
\put(4501,-2311){\makebox(0,0)[lb]{\smash{{\SetFigFont{12}{14.4}{\rmdefault}{\mddefault}{\updefault}{\color[rgb]{0,0,0}\parbox{3.5cm}{\begin{center}Branch point\\(monodromy exchanges sheets)\end{center}}}%
}}}}
\end{picture}%